\documentclass[11pt,a4paper]{article}
\usepackage{jcappub}

\usepackage{bm}
\usepackage[utf8]{inputenc}
\usepackage{amsmath,amsfonts,amssymb}
\usepackage{amsthm}
\usepackage{accents}
\usepackage{enumitem}
\usepackage{graphicx}
\usepackage[usenames,dvipsnames]{xcolor}
\usepackage{hyperref}
\hypersetup{colorlinks=true,
citecolor=ForestGreen, linkcolor=RoyalBlue, urlcolor=Turquoise}
\usepackage{subcaption}
\usepackage{float}
\usepackage{orcidlink}
\newcommand{\oo}{\"{o}}
\newcommand{\ra}{\rightarrow}

\DeclareUnicodeCharacter{2212}{\ensuremath{-}}
\newcommand{\lc}[1]{\overset{\circ}{#1}\vphantom{#1}}

\allowdisplaybreaks

\title{Neutron stars in scalar torsion theories with non minimal coupling}
	
\author[a]{Youcef Kehal\orcidlink{0009-0004-9312-720X},}
\author[a]{Khireddine Nouicer \orcidlink{0000-0001-6110-9833}}
\author[b]{and Hamza Boumaza \orcidlink{0000-0001-8508-7106}}

\affiliation[a]{Laboratory of Theoretical Physics and Department of Physics,\\ Faculty of Exact and Computer Sciences, University of Jijel,\\ B.P. 98, Ouled Aissa, Jijel 18000, Algeria}
\affiliation[b]{Laboratory of Particle Physics and Statistical Physics  (LPPPS),\\ Ecole Normale Supérieure-Kouba,\\ B.P. 92, Vieux Kouba,  Algiers 16050, Algeria
}

\emailAdd{youcef.kehal@univ-jijel.dz}
\emailAdd{khnouicer@univ-jijel.dz}
\emailAdd{boumaza14@yahoo.com}

	\abstract{
		We explore the existence and configurations of static and slowly rotating neutron stars (NSs) within a specific truncation of teleparallel scalar torsion theory. In this model, a scalar field $\phi$ is non-minimally coupled to the torsion scalar as $\xi T \phi^2$, in the presence of the scalar potential $V({\phi}) = −\mu^2 {\phi}^2 /2 + \lambda {\phi}^4 /4$. We establish the hydrostatic equilibrium equations for the static scenario and numerically solve them for both interior and exterior regions, employing appropriate boundary conditions near the center and at a distant location far away from the star's surface. Radial profiles of metric functions and the scalar field, alongside mass-radius diagrams, are plotted, utilizing four different realistic equations of state (EOS). Our results align closely with observational constraints from the GW170817 event, revealing a maximal mass of $2.37 M_\odot$ achieved with the BSk21 EOS for a coupling parameter $\xi = 0.25$. Extending our analysis to encompass slow rotation, we establish the relationship between the star’s moment of inertia and its mass. Furthermore, we explore future observations of NSs utilizing the redshift surface observable. Finally, we demonstrate the validity of the universality relation between the two forms of normalized moment of inertia within teleparallel scalar torsion theory with non minimal coupling.
}
	
	\arxivnumber{2305.12155}

\begin{document}
\maketitle


\section{Introduction}\label{sec_intro}

Beside its mathematical elegance in describing the geometry of spacetime, general relativity (GR) has also been successfully tested in weak gravitational regime. It fits properly with observational data inside the solar and stellar system scales, as it was able to accurately predict a large number of astrophysical phenomena \cite{will2014confrontation}. Nonetheless, the theory seems to be struggling to cope with cosmological facts since that Supernovae Ia data (SN Ia) \cite{riess1998observational,perlmutter1999measurements} and cosmic microwave background (CMB) measurements \cite{spergel2003first} suggest that the universe is currently passing through an accelerated expansion phase which can not be occurred without introducing an exotic form of energy with negative pressure, called dark energy (DE), or by replacing GR with a more complete theory of gravity. Also, on larger scales the existence of hypothetical dark matter (DM) is necessary to fully explain the rotation curves of galaxies and structure formation. All these considerations, along with the challenge of unifying Einstein's theory with quantum mechanics, have spurred researchers to look for alternative theories of gravity.

Efforts in this direction have garnered significant attention, notably following the first detection of gravitational waves (GWs) GW150914 by the LIGO-VIRGO collaboration. This observation resulted from the merger of two binary black holes with masses of approximately $29M_{\odot}$ and $36M_{\odot}$, leading to the formation of a black hole with a mass of $62M_{\odot}$ \cite{abbott2016gw150914}. The event radiated approximately $3.0_{-0.5}^{+0.5}$ solar masses of energy \cite{abbott2016observation}. Subsequently, detections such as GW170817, originating from the coalescence of nearly equal mass binary neutron stars (NSs) with a combined mass of approximately $2.74_{-0.04}^{+0.04}M_{\odot}$ \cite{abbott2017gw170817}, have further propelled research in this area. The electromagnetic counterpart, GW170817A, detected by the Fermi GRB monitor \cite{LIGOScientific:2017zic}, marked the establishment of multi-messenger GW astronomy \cite{metzger2017welcome}. Events of this nature provide invaluable insights into their sources and enable the exploration of physics in regions of high density.

Compact objects like NSs, remnants of stellar collapse, offer unique opportunities to study physics in extreme conditions. With densities reaching approximately $2.8 \times 10^{14}\, \text{g/cm}^3$ and masses typically ranging from $1.4$ to $2M_{\odot}$, NSs serve as ideal astrophysical laboratories for testing modified theories of gravity in strong field regimes \cite{olmo2020stellar}. Their compactness, coupled with the immense gravitational forces at play, provides a rich environment for investigating the fundamental nature of gravity and matter under extreme conditions.

In this context, one straightforward approach to gravity modification involves replacing the Ricci scalar $R$ in the Einstein-Hilbert action with a general function of it, denoted as $f(R)$ \cite{sotiriou2010f}. These theories have received considerable attention, particularly in efforts to elucidate the accelerated expansion of the universe and the inflationary epoch \cite{de2010f}. They have also been extensively explored in the context of NS structure within spherically symmetric and static backgrounds, considering R-squared gravity and Starobinsky models through perturbative \cite{cooney2010neutron,arapouglu2011constraints,orellana2013structure,astashenok2013further,resco2016neutron} or exact and self-consistent  analyses \cite{ganguly2014neutron,yazadjiev2014non,capozziello2016mass,feng2018equation,astashenok2017realistic}.

Moreover, the introduction of an additional scalar degree of freedom  alongside the usual two tensor modes of GR opens up diverse avenues for gravity modification, known as scalar-tensor theories (STT) \cite{quiros2019selected}. This incorporation of scalar fields into the Lagrangian is motivated by the desire to circumvent the assumptions of the Lovelock theorem, often necessitating the use of screening mechanisms to reconcile with local experiments. Numerous classes of STT have been extensively scrutinized to assess their phenomenological implications, especially on highly relativistic objects \cite{babichev2010relativistic,sotani2012slowly,minamitsuji2016relativistic}.

In particular, investigations into NS solutions have been conducted within the framework of Brans-Dicke theories \cite{kase2019neutron}, considering various coupling values between the scalar field and the gravitational sector, and employing massive and self-interacting potentials. Additionally, studies have focused on the effects of a non-minimally coupled scalar field in the Jordan frame on the mass-radius profile of the star, exploring a range of potential choices \cite{arapouglu2019neutron,arapouglu2020neutron,odintsov2022neutron}.

Furthermore, a broader class of STT, including Horndeski theories \cite{kobayashi2019horndeski} and their generalization known as Degenerate Higher-Order Scalar-Tensor (DHOST) theories \cite{langlois2016degenerate}, have analyzed NS configurations \cite{cisterna2016slowly,boumaza2022neutron} within the Hartle-Thorne slow rotation approximation \cite{hartle1967slowly}. This approach, coupled with the consideration of massive scalar fields, introduces more significant deviations from GR due to the finite range of the scalar field determined by its Compton wavelength. Observations beyond this scale outside the star do not tightly constrain the parameters of massive STT, unlike the massless case \cite{yazadjiev2016slowly,staykov2018static}.

In addition to their conventional properties, NSs exhibit novel features in certain theories. These include the emergence of non-trivial scalar hairs, initially discovered by Damour and Esposito-Farèse \cite{PhysRevLett.70.2220} and extensively discussed in subsequent works \cite{salgado1998spontaneous,pani2014slowly,yagi2021neutron,hu2021scalarized,ventagli2021neutron}, the manifestation of spontaneous vectorization in Proca theories \cite{kase2020neutron}, and the potential existence of new types of NSs with topological charges \cite{doneva2020topological}. Moreover, numerous studies have explored universal relations between the normalized moments of inertia ($I$) and the stellar compactness ($C$) of the star \cite{breu2016maximum,staykov2016moment,popchev2019moment}. To expand this understanding, driven by astrophysical phenomena such as recycled pulsars and highly magnetized stars where rotational effects are significant, investigations into NS dynamics can be extended to incorporate the rapidly rotating regime \cite{doneva2013rapidly,yazadjiev2015rapidly}.

Similarly, to elucidate gravity in cosmological and astrophysical systems and to investigate whether including scalar fields and torsion can better describe gravitational phenomena observed in NSs, one can adopt a different perspective by working within the framework of non-Riemannian teleparallel geometry \cite{bahamonde2021teleparallel}. The teleparallel equivalent of General Relativity (TEGR) can be viewed as a gauge theory for the translation group, where the gravitational interaction is not geometrized but is instead represented by the torsion scalar $T$ \cite{aldrovandi2012teleparallel,krvsvsak2019teleparallel}. Analogous to $f(R)$ models, the TEGR Lagrangian can be extended to $f(T)$ theories, or augmented with scalar fields $f(T,\phi)$  \cite{cai2016f,Geng_2011,horvat2015nonminimally}, and even more general constructions such as $f(T,\phi,X,Y)$, which include two additional scalar quantities: the kinetic energy term $X$ and the derivative coupling term $Y$—collectively termed scalar torsion theories \cite{hohmann2018scalar1,hohmann2018scalar2,hohmann2018scalar3}.

Although modified TEGR theories are often criticized for lacking invariance under local Lorentz transformations \cite{li2011f}, this issue has been addressed by reintroducing the spin connection as a dynamical variable alongside the tetrads, thereby establishing the covariant version of TEGR \cite{krvsvsak2016covariant}. Consequently, there has been considerable interest in exploring the impact of $f(T)$ theories on the configuration of stars in static and spherically symmetric spacetime \cite{boehmer2011existence,lin2022realistic,fortes2022solving}, as well as utilizing scalar torsion theories to describe slowly rotating NSs \cite{boumaza2021slowly}. Finally, let us mention that in quantum cosmology, non-zero torsion is regarded as a dynamical degree of freedom and must be considered in the analysis of cosmological models \cite{moniz2010quantum1,moniz2010quantum2}. 

The richer structure of teleparallel scalar-torsion theory (TST) has been demonstrated in the case of non-minimally coupled canonical scalar field to the scalar of torsion in the cosmological context. This richness manifests itself in the absence of a conformal transformation reducing it to a minimally coupled model, unlike in GR \cite{Geng_2011}. Recently, this richness has also been revealed in teleparallel scalar Gauss-Bonnet theory, where a family of scalarized black hole solutions has been found exhibiting richer properties compared to scalar Gauss-Bonnet theory in GR \cite{2023PhRvD.107j4013B}.

In this study, we thoroughly test the richness of teleparallel theory by examining the existence of NS solutions within a specific truncation of TST theories, denoted as $F(T,\phi,X)$. This theory encompasses a non-minimally coupled scalar field to the torsion scalar through the parameter $\xi$. Additionally, we consider a scalar potential of the form $V(\phi)=-\mu^2\phi^2/2 +\lambda \phi^4 /4$. This potential has primarily been investigated alongside a coupling function to matter within the context of the symmetron screening mechanism \cite{hinterbichler2010screening}.

Following the derivation of modified Tolman-Oppenheimer-Volkoff (TOV) equations, we establish that the exterior solution adopts the Schwarzschild-Anti de Sitter (SAdS) form, where the effective cosmological constant arises from a non-vanishing potential $V(\phi_{\infty})$ at spatial infinity. Our primary objective is to construct the mass-radius diagram and juxtapose our findings with available observational data, utilizing four different realistic equations of state (EOS): SLy, BSk19, BSk20, and BSk21. Particularly noteworthy, the GW170817 event has set a constraint on the maximum mass of one of the NS involved in the collision, $M<2.7 M_{\odot}$ \cite{abbott2017gw170817, Margalit_2017}. These results are pivotal in constraining the EOS for matter at extremely high densities, as well as the parameters of any given theory.

Through numerical simulations, we investigate the compatibility of these constraints with our model, yielding a maximum mass of $2.37 M_{\odot}$ for the BSk21 EOS with a coupling parameter $\xi=0.25$. Throughout our inquiry, we aim to elucidate the inertia-mass relations by extending the study into the slow rotation approximation. To this end, we adopt a Weitzenböck regular tetrad to generate an axially symmetric spacetime. We conclude this work by demonstrating the independence of dimensionless and normalized moments of inertia on the equation of state (EOS).

The paper is organized as follows. In Sec. \ref{sec_scalar_torsion}, we establish the groundwork by presenting the general formalism and field equations of TST theories described by a Lagrangian density of the form  $F(T,\phi,X)$, and we discuss the notion of axial symmetry in teleparallel gravity. In Sec. \ref{sec_static}, we derive the TOV equations of hydrostatic equilibrium, along with the initial conditions at the core of the NS, and examine the asymptotic behavior of metric functions, mass, pressure, and scalar field at radial infinity. In Sec. \ref{sec_rotation}, we explore the Hartle-Thorne formalism for slowly rotating NS. In Sec. \ref{sec_profiles}, we discuss the observational consequences of TST theories with non-minimal coupling and compare them with standard results from GR. Conclusions are provided in Sec. \ref{sec_conclusion}.


\section{Scalar torsion theories}\label{sec_scalar_torsion}

\subsection{Action and field equations}\label{subsec_tegr}

This subsection aims to provide a brief overview of the main components of TEGR theory, where the dynamical variables are typically represented by the tetrad fields $e^{a}=e^{a}{}_{\mu }dx^{\mu }$ and a flat spin connection $\omega ^{a}{}_{b}=\omega ^{a}{}_{b\mu }dx^{\mu }$. The tetrads constitute a set of four linearly independent vectors forming an orthonormal basis in the tangent space at each point of the spacetime manifold $(\mathcal{M},g_{\mu \nu })$, while the spin connections are one-forms representing an additional degree of freedom (DoF) associated with inertial effects. It's important to note that throughout this section, Greek letters $(\mu ,\nu ,\rho
...=0,1,2,3)$ label the external spacetime coordinates, whereas Latin letters $(a,b,c...=0,1,2,3)$ label the internal Lorentz coordinates. The metric tensor $g_{\mu \nu }$ can be constructed via the relation:
\begin{equation}\label{eq_metric_tensor}
	g_{\mu \nu }=\eta _{ab} e^a{}_{\mu}e^b{}_{\nu} \,,
\end{equation}
where $\eta_{ab}=\mathrm{diag}(-1,1,1,1)$ represents the Minkowski metric associated with the tangent space. The tetrad components are normalized as $ e^a{}_{\nu}e_a{}^{\mu}=\delta^{\mu}_{\nu} $ and $ e^a{}_{\mu}e_b{}^{\mu}=\delta^{a}_{b} $. Under local Lorentz transformations (LLTs) $\lambda ^{a}{}_{b}(x)$, the metric tensor (\ref{eq_metric_tensor}) remains invariant, and for a fully covariant approach, both the tetrads and spin connections must transform simultaneously. \cite{bahamonde2021teleparallel}
\begin{equation}\label{eq_LLT}
	e^{\prime a}{}_{\mu}=\lambda ^{a}{}_{b} \, e^{b}{}_{\mu } \, , \quad \omega ^{\prime a}{}_{b\mu }=\Lambda ^{a}{}_{c} \, \omega ^{c}{}_{d\mu}(\Lambda ^{-1})^{d}{}_{b}+\Lambda ^{a}{}_{c} \, \partial _{\mu }(\Lambda
	^{-1})^{c}{}_{b} \, .
\end{equation}
The spin connection plays a crucial role in compensating inertial effects \cite{krvsvsak2019teleparallel}, and its presence is essential to maintain the invariance under local Lorentz transformations (LLTs) and ensure the frame-independence of the entire theory \cite{krvsvsak2016covariant}. Locally, it is advantageous to work within a purely inertial gauge where the spin connection vanishes identically. This gauge choice is often referred to as the Weitzenböck gauge, expressed as:
\begin{equation}
\omega ^{a}{}_{b\mu }=\Lambda ^{a}{}_{c}  \partial_{\mu }\Lambda_{b}{}^{c} .
\end{equation}
In contrast to the torsion-less Levi-Civita connection $ {\overset{\circ}{\Gamma}}\vphantom{\Gamma}^{\rho}{}_{\mu \nu} $ of GR, where the gravitational interaction is geometrized by curvature \cite{aldrovandi2012teleparallel}, the formulation of TEGR adopts the so-called Weitzenböck connection $\Gamma ^{\rho }{}_{\mu \nu }$. This connection is the unique linear affine connection that satisfies the metric compatibility condition, possesses torsion, and results in vanishing curvature. Therefore, the field strength of gravity is represented by the torsion tensor
\begin{equation}\label{eq_tor}
	T^{\rho }{}_{\mu \nu }=-2\Gamma ^{\rho }{}_{[\mu \nu ]}=e_{a}{}^{\rho
	}\left( \partial _{\mu }e^{a}{}_{\nu }-\partial _{\nu }e^{a}{}_{\mu }+\omega
	^{a}{}_{b\mu }e^{b}{}_{\nu }-\omega ^{a}{}_{b\nu }e^{b}{}_{\mu }\right) \, ,
\end{equation}
which  is invariant under LLTs (\ref{eq_LLT}) and diffeomorphisms. The tensor that relates the Levi-Civita and Weitzenb\oo ck connections,  the contortion tensor,  is given by Ricci theorem
\begin{equation} \label{eq_kot}
K^{\rho }{}_{\mu \nu }=\Gamma ^{\rho }{}_{\mu \nu }-\overset{\circ }{\Gamma }\vphantom{\Gamma}^{\rho }{}_{\mu \nu }  
		 =\frac{1}{2}\left( T_{\mu \ \nu }^{\ \rho }+T_{\nu \ \mu }^{\ \rho }-T_{\
			\mu \nu }^{\rho }\right) \, .
\end{equation}
It is also useful to define the superpotential tensor as follows:
\begin{equation}\label{eq_sup}
	S_{\rho }{}^{\mu \nu }=K^{\mu \nu }{}_{\rho }-\delta _{\rho }^{\mu
	}T_{\sigma }{}^{\sigma \nu }+\delta _{\rho }^{\nu }T_{\sigma }{}^{\sigma \mu
	} \, .
\end{equation}
Therefore, the torsion scalar can be expressed in compact form as:
\begin{equation}\label{eq_tor_scalar}
	T=\frac{1}{2}S_{\rho }{}^{\mu \nu }T^{\rho }{}_{\mu \nu } \, .
\end{equation}
Indeed, the Lagrangian of TEGR is constructed from a linear form of the torsion scalar $T$, which differs from the Ricci scalar $ \overset{\circ}{R} $ only by a total divergence term $B$, defined as follows:
\begin{equation}\label{eq_boundary}
			\overset{\circ }{R}=-T+B  \\
		 =-T+\frac{2}{e}\partial _{\mu }\left( eT_{\lambda }{}^{\lambda \mu
		}\right),
\end{equation}
where $e=\textrm{det}(e^a_{\,\,\mu})=\sqrt{-g}$. The relation (\ref{eq_boundary}) demonstrates that the Lagrangians of TEGR and GR are equivalent up to a surface term. This equivalence ensures that their corresponding field equations, and thus all the phenomenological predictions at astrophysical and cosmological levels, remain the same. This equivalence is represented by the following identity:
\begin{equation}\label{eq_identity}
	G_{\alpha \beta }=\overset{\circ }{\nabla }\vphantom{\nabla}_{\rho }S_{\beta
		\alpha }{}^{\rho }-\left(K^\rho{}_{\beta \nu}+T^\rho{}_{\beta \nu}\right)S_{\rho \alpha}{}^{\nu}+\frac{1}{2}g_{\alpha \beta}T ,
\end{equation}
where $ G_{\alpha \beta} $ is the Einstein tensor, and the contorsion tensor (\ref{eq_kot}) can be expressed in terms of the spin connections as $ K^{b}{}_{a\mu}=\omega ^{b}{}_{a\mu }-\overset{\circ }{\omega }\vphantom{\omega}
^{b}{}_{a\mu }$. Additionally, the Levi-Civita covariant derivative is defined as:
\begin{equation}
	\overset{\circ }{\nabla }\vphantom{\nabla}_{\nu }S_{a}{}^{\mu \nu
		}=e^{-1}\partial _{\nu }\left( eS_{a}{}^{\mu \nu }\right) -{\overset{
			\circ }{\omega }\vphantom{\omega}^{b}{}_{a\nu }}S_{b}{}^{\mu \nu}  .
\end{equation}
The TEGR Lagrangian can be extended by adding scalar fields $\phi$ and their kinetic energy term $X = -\frac{1}{2}g^{\mu \nu}\phi_{,\mu}\phi_{,\nu}$ in the Lagrangian density. This changes the action as follows:
\begin{equation}\label{eq_action}
	\mathcal{S}=-\frac{1}{2}\int d^{4}x\,e\, F\left( T,\phi,X \right) +\int d^{4}x\,e\,\mathcal{L}_{m},
\end{equation}
where $F$ is a function of all its arguments, and $\mathcal{L}_{m}$ denotes the matter Lagrangian density. We assume that the matter fields are minimally coupled to the scalar field but are not allowed to couple directly with the spin connection \cite{hohmann2018scalar1}. The energy-momentum tensor is defined as:
\begin{equation}\label{eq_energy_momentum}
	\Theta _{\mu \nu }=e^{a}{}_{\mu }\left( \frac{1}{e}\frac{\delta \left( e
		\mathcal{L}_{m}\right) }{\delta e^{a}{}_{\lambda }}\right) g_{\nu \lambda
	}=e^{a}{}_{\mu }\Theta _{a}{}^{\lambda }g_{\nu \lambda }  .
\end{equation}
The variations of the action (\ref{eq_action}) with respect to the tetrad fields yield:
\begin{equation}
	-\frac{1}{2} F e_a{}^{\mu}+{\overset{\circ}{\nabla}}_{\nu}\left({F_T S_a{}^{\mu \nu}}\right)-F_T \left(T^{\sigma}{}_{a \nu}+K^{\sigma}{}_{a \nu}\right)S_{\sigma}{}^{\mu \nu}+\frac{1}{2}F_X g^{\mu \nu}e_a{}^{\rho} \phi_{,\nu} \phi_{,\rho}=\Theta _{a}{}^{\mu } \, .
\end{equation}
Contracting with $ e^{a}{}_{\beta }\, g_{\mu \alpha } $ and using  (\ref{eq_identity}), we  express the field equations in covariant form:
\begin{equation}\label{eq_tetFEQ}
	{G}_{\alpha \beta }F_{T}+\frac{1}{2}g_{\alpha \beta }\left( F-F_{T}T\right) +S_{\beta \alpha }{}^{\rho }{}\overset{\circ}{\nabla }_{\rho }F_{T}+\frac{1}{2}F_{X}\phi
	_{,\alpha }\phi_{,\beta } = \Theta_{\alpha \beta },
\end{equation}
where $ F_{A}\equiv \partial F/\partial A$ and $ A=T,X,\phi $. Here $1/F_T$ is an effective gravitational coupling. The Eqs.(\ref{eq_tetFEQ}) belong to the extended theories of gravitation (ETG) \cite{Wojnar_2016}. 

The antisymmetric part of (\ref{eq_tetFEQ}) coincides with the variations of the action (\ref{eq_action}) with respect to the spin connection \cite{hohmann2018scalar2}:
\begin{equation}\label{eq_spiFEQ}
	\left(S_{\alpha \beta}{}^{\rho}-S_{\beta \alpha}{}^{\rho}\right)\overset{\circ}{\nabla}_{\rho}F_T = 0 \, .
\end{equation}
Varying  (\ref{eq_action}) with respect to the scalar field $ \phi $ leads to the Klein-Gordon equation:
\begin{equation}\label{eq_scaFEQ}
	g^{\mu\nu}\overset{\circ}{\nabla}_{\mu}\left(F_X \phi_{,\nu}\right)+F_{\phi}=0.
\end{equation}
Finally, applying the covariant divergence to \eqref{eq_tetFEQ}, we obtain:
\begin{equation}
\lc{\nabla}^{\alpha} \Theta_{\alpha \beta }=\frac{1}{2}\lc{\nabla}^{\alpha}F_{T}S_{\mu\nu\alpha}K^{\mu\nu}{}_{\beta}\equiv 0,
\end{equation}
by virtue of antisymmetry of the contorsion tensor and symmetry of the superpotential tensor (as given by Eq. \eqref{eq_spiFEQ} in their two first indices. Hence, the matter energy-momentum tensor is covariantly conserved \cite{Hohmann_2018}. 
\subsection{Axial symmetric spacetime}\label{subsec_axial}

After introducing the fundamental concepts of teleparallel gravity and presenting the field equations of the TST theory with the Lagrangian density $F(T,\phi,X)$, we  provide now a brief overview of symmetries within the teleparallel geometric framework \cite{hohmann2019modified}. Subsequently, we proceed to derive the Weitzenböck tetrad that yields an axially symmetric spacetime \cite{bahamonde2021exploring}, characterized by a non-diagonal metric tensor with one cross term.

The teleparallel geometry, denoted as $(\mathcal{M},\bm{e},\bm{\omega})$, is invariant under the action $\Phi $ of a Lie group $G$ on a spacetime manifold $\mathcal{M}$ if and only if the metric and the affine connection are invariant under the diffeomorphisms $\Phi_{u}:\mathcal{M}\rightarrow \mathcal{M}$ such that $u\in G$. For convenience, when discussing teleparallel symmetries, we focus solely on the infinitesimal symmetries where the diffeomorphisms $\Phi_{u}$ are generated by a set of vector fields $X_{\varsigma }$ that preserve the geometric structure of the spacetime $\mathcal{M}$. This, in turn, requires that the Lie derivative of the metric tensor as well as the teleparallel affine connection vanishes in the direction of $X_{\varsigma }$:
\begin{equation}\label{eq_lie_derivative}
	\mathcal{L}_{X_{\varsigma }}g_{\mu \nu }=0,\quad \mathcal{L}_{X_{\varsigma
	}}\Gamma ^{\lambda }{}_{\mu \nu }=0.
\end{equation}
A teleparallel geometry is symmetric only if there exists a local Lie algebra homomorphism $\bm{\lambda}$ mapping the symmetry group into the Lorentz group $SO(1,3)$, as defined by \cite{hohmann2019modified}:
\begin{equation}
	\lambda _{\varsigma }(x)=\frac{d}{dt}\Lambda _{\varsigma (t)}(x),
\end{equation}
where $\Lambda :G\times \mathcal{M}\ra SO(1,3)$ is the corresponding local Lie group homomorphism. The requirement of invariance, i.e the vanishing Lie derivatives (\ref{eq_lie_derivative}),  implies that the Lie derivative of the tetrads and the spin connections must vanish:
\begin{equation}\label{eq_tele_killing}
	\mathcal{L}_{X_{\varsigma }}e^{a}{}_{\mu }=-\lambda ^{a}{}_{b}e^{b}{}_{\mu
	} \,, \quad \mathcal{L}_{X_{\varsigma }}\omega ^{a}{}_{b\mu }=\xi_{\mu
	}\lambda ^{a}{}_{b}=\partial _{\mu }\lambda ^{a}{}_{b}+\omega ^{a}{}_{c\mu
	}\lambda ^{c}{}_{b}-\omega ^{c}{}_{b\mu }\lambda ^{a}{}_{c} \, .
\end{equation}
These are the teleparallel Killing equations. Clearly, it is much easier to solve them in the Weitzenböck gauge by selecting a particular choice of $\Lambda ^{a}{}_{b}$ where the spin connections vanish. By doing so, the second condition of (\ref{eq_tele_killing}) implies that the local Lie algebra homomorphism $\lambda $ is spacetime-independent, and hence the symmetric tetrads satisfying the first condition of (\ref{eq_tele_killing}) are related by a global Lorentz transformation \cite{hohmann2019modified}. It can be stated that the symmetries of the tetrad-spin connection pair guarantee that the torsion tensor also satisfies these symmetries. We require that the symmetry conditions are also fulfilled by the scalar fields \cite{bahamonde2021exploring}:
\begin{equation}
	\mathcal{L}_{X_{\varsigma }}T^{\rho }{}_{\mu \nu }=0 \, , \quad \mathcal{L}
	_{X_{\varsigma }}\phi =0 \, .
\end{equation}
In spherical coordinates $(t,r,\theta ,{\varphi})$, the single Killing vector field that forms the $so(2)$ algebra and generates the $SO(2)$ group is $X_{z}=\partial _{{\varphi} }$, which generates a rotation $R({\varphi} )\in SO(2)$ around the polar axis. Within the Weitzenböck gauge, one needs to determine the Lie algebra homomorphism $\lambda :g\rightarrow so(1,3)$, ensuring that $\lambda $ is constant over the spacetime. While multiple choices are possible, we restrict ourselves to a nontrivial canonical one that yields the regular branch of the Weitzenböck tetrads, whose components take the form \cite{hohmann2019modified}:
\begin{equation}\label{eq_general_tetrad}
	e^{0}{}_{\mu }=C^{0}{}_{\mu }\, ,\quad e^{1}{}_{\mu }=C^{1}{}_{\mu }\cos
	{\varphi} -C^{2}{}_{\mu }\sin {\varphi} \, ,\quad e^{2}{}_{\mu }=C^{1}{}_{\mu }\sin
	{\varphi} +C^{2}{}_{\mu }\cos {\varphi} \, ,\quad e^{3}{}_{\mu }=C^{3}{}_{\mu } \, .
\end{equation}
This branch exhibits a proper limit to spherically symmetric tetrads, where the functions $C^{a}{}_{\mu}$ become dependent on the coordinates $(r,\theta )$ after assuming stationarity. Generally, substituting the tetrad (\ref{eq_general_tetrad}) into (\ref{eq_metric_tensor}) results in a non-circular metric with more than one off-diagonal component. To achieve an axially symmetric spacetime with only one cross term, we impose a particular coordinate gauge to eliminate all off-diagonal components of the metric tensor except for the $dt,d\tilde{\phi} $ component \cite{bahamonde2021exploring}:
\begin{eqnarray}\label{eq_tetrad_constraints}
	&&C^{0}{}_{1}=C^{1}{}_{0}=C^{0}{}_{2}=C^{2}{}_{0}=C^{1}{}_{3}=C^{2}{}_{1}=C^{2}{}_{2}=C^{3}{}_{0}=C^{3}{}_{3}=0 \, \nonumber \\   && \qquad \qquad \qquad C^{2}{}_{3}=-C^{3}{}_{2}\, , \quad C^{3}{}_{1}=-
	\frac{C^{1}{}_{1}C^{1}{}_{2}}{C^{2}{}_{3}} \, ,
\end{eqnarray}
and we are only left with $5$ functions to determine. The components of the axial symmetric metric for the reduced tetrad are then given by \cite{bahamonde2021exploring}
\begin{eqnarray}
	&&g_{tt}=-\left( C^{0}{}_{0}\right) ^{2},\, g_{rr}=\left( C^{1}{}_{1}\right)
	^{2}\left[ \left( \frac{C^{1}{}_{2}}{C^{2}{}_{3}}\right) ^{2}+1\right],\, g_{\theta \theta}= \left( C^{1}{}_{2}\right) ^{2}+\left( C^{2}{}_{3}\right) ^{2} \nonumber \\
	 && \quad \quad \quad \quad \quad g_{\tilde{\phi} \tilde{\phi}}= \left( C^{2}{}_{3}\right) ^{2}-\left( C^{0}{}_{3}\right) ^{2}%
,\, g_{t \tilde{\phi}}=-C^{0}{}_{0}C^{0}{}_{3} \, .
\end{eqnarray}
The presence of the free function $C^{0}{}_{3}$ induces rotation around the polar axis. However, in static and spherically symmetric spacetimes, $C^{0}{}_{3}$ is expected to vanish.

\section{Static and spherically symmetric background}\label{sec_static}

\subsection{Modified Tolman-Oppenheimer-Volkoff equations}\label{subsec_tov}

In this section we derive the equations of
hydrostatic equilibrium that are solutions of the 
field equations of TST theory derived in the previous section. For this
we shall assume a static, spherically symmetric form for
the metric given by:
\begin{equation}\label{eq_static_metric}
ds^{2}=-f(r)dt^{2}+h(r)dr^{2}+r^{2}(d\theta ^{2}+\sin^{2}\theta d{\varphi}
^{2}).
\end{equation}
This spacetime characterizes regions both inside and outside of compact astronomical objects such as NSs, black holes, and white dwarfs in Schwarzschild coordinates (for a comprehensive overview, see \cite{glendenning1997compact}).

Among an infinity possible tetrads that can recover the metric (\ref{eq_static_metric}), the
simplest is the diagonal  one given by:
\begin{equation} \label{eq_diag_tet}
	\tilde{e}^{a}{}_{\mu }=\mbox{diag}(\sqrt{f(r)},\sqrt{h(r)},r,r\mbox{sin}
	\theta ) \, .
\end{equation}
The literature often characterizes this tetrad as a "bad" choice for spherically symmetric solutions in $f(T)$ gravity \cite{tamanini2012good}, primarily due to its introduction of extra off-diagonal components in the field equations. Consequently, this constrains the functional form of $f(T)$, effectively reducing the modified theory to TEGR theory \cite{boehmer2011existence}. However, as demonstrated in \cite{krvsvsak2016covariant}, this issue arises from the old pure-tetrad formulation of teleparallel gravity, which assumes a vanishing spin connection. In TEGR, the spin connection enters the action as a total divergence term, making any arbitrary choice of the spin connection inconsequential to the TEGR field equations, leaving them invariant under Local Lorentz Transformations (LLTs). Nonetheless, overlooking the spin connection's role presents a delicate problem when extending the action to a non-linear general function of $T$. Consequently, the standard $f(T)$ gravity formulation evidently suffers from Lorentz invariance violation. The resolution to this problem lies in adopting a covariant version of $f(T)$ \cite{krvsvsak2016covariant} and computing the associated spin connections for each tetrad using the switch-off gravity technique outlined in \cite{krvsvsak2015spin}. The non-vanishing components of the spin connection for the diagonal tetrad \eqref{eq_diag_tet} are then given by:

\begin{equation}\label{eq_spin}
	\omega _{\ 2\theta }^{1}=-1\, ,\quad \omega _{\ 3\tilde{\phi} }^{
			1}=-\sin \theta \, , \quad \omega _{\ 3\tilde{\phi} }^{2}=-\cos \theta \, .
\end{equation}
Through an ansatz choice based on symmetry principles \cite{hohmann2019modified} and discussed above in subsection (\ref{subsec_axial}), one can also reproduce the metric (\ref{eq_static_metric}) by using a rotated tetrad $e^{a}{}_{\mu }$ related to the diagonal one (\ref{eq_diag_tet}) by a Lorentz transformation matrix (\ref{eq_LLT}) such that
\begin{equation}\label{eq_tetrad_coef1}
	C^{0}{}_{0}=\sqrt{f(r)} \, , \quad C^{1}{}_{1}=\sqrt{h(r)}\sin \theta \, , \quad
	C^{1}{}_{2}=r\cos \theta \, , \quad C^{2}{}_{3}=r\sin \theta \, , \quad
	C^{0}{}_{3}=0 \, .
\end{equation} 
Inserting the functions \eqref{eq_tetrad_coef1} into \eqref{eq_general_tetrad} and utilizing the constraints \eqref{eq_tetrad_constraints}, the non-diagonal tetrad assumes the following form:
\begin{equation}\label{eq_tetrad_static}
	e^{a}{}_{\mu}=\left( 
	\begin{array}{cccc}
		\sqrt{f(r)} & 0 & 0 & 0 \\ 
		0 & \sqrt{h(r)}\sin (\theta )\cos ({\varphi} ) & r\cos (\theta )\cos ({\varphi} ) & 
		-r\sin (\theta )\sin ({\varphi} ) \\ 
		0 & \sqrt{h(r)}\sin (\theta )\sin ({\varphi} ) & r\cos (\theta )\sin ({\phi} ) & 
		r\sin (\theta )\cos ({\varphi} ) \\ 
		0 & \sqrt{h(r)}\cos (\theta ) & -r\sin (\theta ) & 0 
	\end{array}
	\right) \, .
\end{equation}
All components of the spin connection associated with this tetrad are null, satisfying the antisymmetric field equation \eqref{eq_spiFEQ}. Therefore, substituting the tetrad \eqref{eq_tetrad_static} and the vanishing spin connection, or the diagonal tetrad \eqref{eq_diag_tet} with their associated spin connections \eqref{eq_spin} to calculate the torsion scalar \eqref{eq_tor_scalar} or the boundary term in \eqref{eq_boundary}, will lead to the same result:
\begin{eqnarray}\label{eq_torsion_scalar}
	T &=&\frac{2\left( \sqrt{h}-1\right) \left[ rf^{\prime }+f\left( 1-\sqrt{h}
		\right) \right] }{r^{2}fh} \, , \label{eq_torsion_calar} \\
	B &=&\frac{4f^{2}\left( 2h^{3/2}+rh^{\prime }-2h\right) +r^{2}f^{\prime
			2}h+r f\left[ (4f^{\prime }h^{3/2}+rf^{\prime }h^{\prime }-2h\left( rf^{\prime
			\prime }+4f^{\prime }\right) \right] }{2r^{2}f^{2}h^{2}} \, . \label{eq_boundary_term}
\end{eqnarray}
Using any tetrad with its corresponding spin connection and the torsion scalar \eqref{eq_torsion_calar} in the field equations \eqref{eq_tetFEQ}, we  derive the $"tt",\, "rr",\, "\theta\theta"$ components of the field equations:
\begin{eqnarray}
	\rho &=& -F+\left[\frac{2}{rh }\left(\left(\sqrt{h}-1\right)\frac{f'}{f}+\frac{h'}{h}\right)+\frac{4}{r^{2}h }\left(\sqrt{h}-1\right)\right]F_{T}+\frac{4}{ r h} \left(\sqrt{h}-1\right)F'_{T} \, , \label{eq_density} \\
	P &=& F+ \left[-\frac{2}{r h }\left(\sqrt{h}-2\right)\frac{f'}{f}-\frac{4}{r^{2}h}\left(\sqrt{h}-1\right)\right]F_{T}+\frac{\phi'^{2}}{h}F_{X} \, , \label{eq_pressure} \\
	P &=& F+\left[\frac{2\left(\sqrt{h}-1\right)^{2}}{r^{2}h}-\frac{f'}{f}\left(\frac{r h f'+f\left(-6h+4h^{3/2}+r h'\right)}{2r f h^{2}}\right)-\frac{h'}{r h^{2}}+\frac{f''}{f h}\right]F_{T} \nonumber \\
	& &-\left[\frac{2f\left(\sqrt{h}-1\right)-rf'}{r f h}\right]F'_{T} \, ,\label{eq_pressure2}
\end{eqnarray}
Similarly, the scalar field equation (\ref{eq_scaFEQ}) yields to
\begin{equation}\label{eq_phi2}
	\phi''+\left(\frac{2}{r}+\frac{1}{2}\left(\frac{f'}{f}-\frac{h'}{h}\right)+\frac{1}{2}\frac{F'_{X}}{F_{X}}\right)\phi'+ \frac{h F_{\phi}}{F_{X}}=0 \, .
\end{equation}
For the matter sector, we model the stellar matter within the NS as a perfect fluid. Therefore, the energy momentum tensor is:
\begin{equation}\label{eq_perfect_fluid}
	\Theta _{\mu \nu }=(\rho +P)u_{\mu }u_{\nu }+Pg_{\mu \nu } \, ,
\end{equation}
where $\rho$, $P$ and $u^{\mu }$ are energy density,  pressure, and fluid's four velocity,  respectively. It's important to note that the fluid is at rest, so the normalization condition $u^{\mu }u_{\mu}=-1$ implies $u^{\mu }=(f(r)^{-1/2},0,0,0)$. The conservation equations $\nabla ^{\mu }\Theta _{\mu \nu }=0$, lead to Euler’s equation:
\begin{equation}\label{eq_continuity}
	P^{\prime }+\frac{f^{\prime }}{2f}(\rho +P)=0 \, .
\end{equation}

Throughout this paper, we propose to study a specific truncation of TST theories described by the Lagrangian density given by 
\begin{equation}\label{eq_model}
		F(T,\phi,X) = \left(\kappa + \xi  \phi^{2}\right)T-2 X +2 V(\phi)\, .
\end{equation}
The action is expressed in the Jordan frame with the non-minimal coupling function $\Omega(\phi )=1+(\xi/\kappa) \phi ^{2}$, where $\xi$ denotes the coupling parameter, $\kappa =1/8\pi G$, and $V(\phi)$ is a scalar potential. The Lagrangian density (\ref{eq_model}) belongs to a broader class of TST theories \cite{hohmann2018scalar1,hohmann2018scalar2,hohmann2018scalar3}, and its functional form bears similarity to the non-minimal quintessence model in TEGR \cite{Geng_2011,D_Agostino_2018}. If $\xi=0$, we recover the GR quintessence model.

In Eqs. (\ref{eq_density})-(\ref{eq_pressure2}), the metric $f(r)$ and its derivatives appear as $f'/f$ and $f''/f$. It proves advantageous, for later use, to utilize the variable $\tilde{f}(r) = f(r)/f(0)$. Then, the field equations for the model (\ref{eq_model}) become:
\begin{eqnarray}
	\frac{\tilde{f}^{\prime }}{\tilde{f}} &=&\frac{h-1}{r}+\frac{
		2r^{2}h(P-V)+r^{2}\phi ^{\prime 2}}{2r\left( \kappa +\xi \phi
		^{2}\right) }\, ,  \label{eq_tov1} \\
	\frac{h^{\prime }}{h} &=&\frac{1-h}{r}+\frac{2r^{2}h(\rho +V)-8\xi \left( \sqrt{h}-1\right) r\phi \phi ^{\prime}+r^{2}\phi ^{\prime 2}}{2r\left( \kappa +\xi \phi ^{2}\right) }\,.
	\label{eq_tov2}
\end{eqnarray}
In the absence of the scalar field these equations reduce  to GR TOV equations \cite{hartle1967slowly}.

Expressed  in terms of (\ref{eq_tov1}), the torsion scalar  reads as:
\begin{equation}\label{eq_tors}
	T=(\sqrt{h}-1)\left[\frac{2(h-\sqrt{h})}{r^{2}h}+\frac{2h\left(P-V\right)+\phi'^{2}}{h\left(\kappa+\xi \phi^{2}\right)}\right] \, .
\end{equation}
Now we substitute (\ref{eq_tov1}) and (\ref{eq_tov2}) into (\ref{eq_phi2}) to get:
\begin{eqnarray}\label{eq_tov3}
	\phi''+\left(\frac{2}{r}-\frac{h'}{2h}+\frac{\tilde{f'}}{2\tilde{f}}\right)-h\left(\xi T \phi + V_{\phi }\right)=0  \, ,
\end{eqnarray}
where $V_{\phi }\equiv \partial V/\partial \phi$.

On the other hand, Euler's equation (\ref{eq_continuity}) becomes:
\begin{equation}\label{eq_tov4}
	P'=-\left( \rho +P\right) \left(\frac{h-1}{2r}+\frac{2rh\left(P-V\right)+r\phi'^{2}}{4\left(\kappa +\xi \phi^{2}\right)}\right) \, .
\end{equation}
Now, we shall define the Arnowitt-Deser-Misner (ADM) mass of the star
by  $M=m_{\infty}$,  where $m_{\infty}$ is the mass function $m(r)= 4 \pi \kappa r \left(1-h^{-1}\right)  $ evaluated at radial infinity. With non minimally coupled scalar field we show that  the mass is governed by \cite{salgado1998spontaneous}:
\begin{equation}\label{eq_mass}
	m'(r)=4\pi\kappa r^2 E(r) \, ,
\end{equation}
where
\begin{equation}\label{eq_mass2}
	E(r)=\frac{1}{\kappa+\xi \phi^{2}}\left[\rho+V(\phi)+\frac{\phi'^{2}}{2h}+\frac{4\xi}{rh}\left(1-\sqrt{h}\right)\phi \phi'\right].
\end{equation}
The set of Eqs. (\ref{eq_tov1}), (\ref{eq_tov2}), (\ref{eq_tov3}), (\ref{eq_tov4}) and (\ref{eq_mass}) describe the  structure of the star and represent the modified TOV equations of hydrostatic equilibrium in static and spherically symmetric background for a canonical scalar field $ \phi $ non-minimally coupled to the torsion scalar $ T $  with  a potential $ V(\phi) $. 

\subsection{Boundary Conditions and external vacuum solution}\label{subsec_boundary}

To determine the initial conditions at the star center $r=0$ and the behavior at spatial infinity for the model $F(T,\phi,X)$ given by \eqref{eq_model}, we need to analyze the behavior of $f,\,h,\,\phi$ and $P$ near $r=0$ and at spatial infinity. 

At $r=0$, we perform Taylor expansions:
\begin{equation}\label{eq_series}
\tilde{f}(r)=1+\sum_{n=2}^{\infty }f_{n}r^{n}, \quad
h(r)=h_{c}+\sum_{n=2}^{\infty }h_{n}r^{n}, \quad
\phi (r)=\phi_
{c}+\sum_{n=2}^{\infty }\phi_{n}r^{n}, \quad
P(r)=P_{c}+\sum_{n=2}^{\infty }P_{n}r^{n}.
\end{equation}
We also expand the scalar potential as:
\begin{equation}
V(\phi)=V(\phi_{c})+\sum_{n=1}^{\infty }\frac{1}{n!}\frac{d^{n}V}{d\phi
^{n}}\biggl|_{\phi =\phi _{c}}(\phi -\phi _{c})^{n}.
\end{equation}
Here $f_{n}$, $h_{c}$, $h_{n}$, $\phi_{c}$, $\phi_{n}$, $P_{c}$, $P_{n}$ are constants, and the lowercase index $"c"$ designates the center of the star. The value of $\phi_c$ is not arbitrary but must be fixed by the value of the field at spatial infinity. To avoid singularities and ensure geometrical regularity of the star at the center, we impose the following boundary conditions:
\begin{equation}\label{eq_regularity}
f'(r_c)=0,\,\quad h'(r_c)=0,\,\quad \phi'(r_c)=0,\,\quad P'(r_c)=0.
\end{equation}
Substituting into the differential equations (\ref{eq_density})-(\ref{eq_pressure}) and (\ref{eq_scaFEQ})-(\ref{eq_continuity}), and equating coefficients of equal powers in $r$, we obtain at the star's center:
\begin{eqnarray}
\tilde{f}(r)&=& 1+\frac{\left(\rho_{c}+3P_{c}-2V(\phi_{c})\right)}{6\left(\kappa+\xi \phi_{c}^2\right)}r^2 +\mathcal{O}(r^4) ,  \label{eq_initial1} \\
h(r)&=&h_{c}+\frac{\left(\rho_{c}+V(\phi_{c})\right)}{3\left(\kappa+\xi \phi_{c}^2\right)}r^2 +\mathcal{O}(r^4) ,  \label{eq_initial2}\\
\phi(r)&=&\phi_{c}+\frac{1}{6}V_{\phi}(\phi_{c})r^2 +\mathcal{O}(r^4) , \label{eq_initial3}\\
P(r)&=&P_{c}-\frac{\left(\rho+P\right)\left(\rho_{c}+3P_{c}-2V(\phi_{c})\right)}{12\left(\kappa+\xi \phi^{2}\right)}r^2 +\mathcal{O}(r^4). \label{eq_initial4}
\end{eqnarray}
Here $\rho_c$ is the central density of the NS, and it is a free parameter that effectively determines the NS mass.
It's worth noting that $V(\phi)$ and the coupling parameter $\xi$ explicitly appear in the metric functions and pressure, while the scalar field $\phi$ is only affected by the gradient of the potential. Similar results have been obtained for NSs in $f(R)$-gravity and ST theories \cite{kase2019neutron}.

Finally, utilizing the mass function \eqref{eq_mass}, we obtain:
\begin{equation}
m(r)=\frac{2\pi \kappa \left( \rho _{c}+V\left( \phi_ {c}\right)
\right) }{\kappa +\xi \phi_ {c}^{2}}r^3 +\mathcal{O}(r^4).\label{eq_adm_general}
\end{equation}
Here, the NS mass at the center is fixed to zero. It's worth noting from (\ref{eq_adm_general}) that the NS mass depends on $ r^3 $, implying $h_{c}=h(r=0)=1$. Consequently, from (\ref{eq_tors}), it's straightforward to conclude that $T(r=0)=0$. The central value of the scalar field $ \phi_{c} $ will be determined using a shooting method to satisfy the boundary condition at spatial infinity, $ \phi(r\rightarrow{\infty})=\phi_{\infty}$.

Outside the star, ensuring asymptotic flatness is crucial in scalar-tensor theories (STT) \cite{kase2019neutron,arapouglu2019neutron,arapouglu2020neutron}. This requirement mandates a vanishing potential as the scalar field approaches a constant asymptotic value $\phi_{\infty}$. Such a condition guarantees that the gravitational field of the NS has a finite range, thus constraining its influence on neighboring objects. Nonetheless, for the subsequent analysis, we assume the NS is enveloped by a cosmological background, with the caveat that the physical properties of the NS remain unaffected by it \cite{sotiriou2012}. Consequently, when substituting \eqref{eq_model} into \eqref{eq_scaFEQ}, the scalar field at spatial infinity satisfies the following equation:
\begin{equation}\label{eq_box}
\Box \phi-V_{\text{eff},\phi}\bigl|_{\phi =\phi_{\infty}}=0,
\end{equation}
where the effective scalar potential is given by:
\begin{equation}\label{eq_pot_eff}
V_{\text{eff}}(\phi)=\frac{1}{2}\xi T \phi^2+V(\phi).
\end{equation}
At spatial infinity,  Eqs. (\ref{eq_tov1}-\ref{eq_tov2}) reduce to:
\begin{equation}
	\frac{\tilde{f}'}{\tilde{{f}}}=-\frac{h'}{h}=-\frac{1-h}{r}-\frac{h r V(\phi_{\infty})}{\kappa +\xi \phi_{\infty}^2},
\end{equation}
whose solutions are simply given by:
\begin{equation}\label{eq_external}
	\tilde{f}(r)=1-\frac{2 G M}{r}-\frac{\Lambda_{\text{eff}}}{3}r^{2},   \,
	h(r)=\frac{1}{1-\frac{2 G M}{r}-\frac{\Lambda_{\text{eff}}}{3}r^{2}},
\end{equation}
where
\begin{equation}
\Lambda_{\text{eff}}=\frac{V(\phi_{\infty})}{\kappa+\xi \phi_{\infty}^{2}}.
\end{equation}
The constant of integration $ M $ is actually the ADM mass   of the star, and $ \Lambda_{\text{eff}} $ is an effective cosmological constant. The solutions (\ref{eq_external}) imply that outside the star we have a Schwarzschild Anti de Sitter (SAdS) space (for negative $ V(\phi_{\infty}) $) or a Schwarzschild de Sitter (SdS) space (for positive $ V(\phi_{\infty}) $). The assumption that $ \Lambda_{\text{eff}}  \ll 1 $, and considering  the weak field approximation $ {M}/r \ll 1 $ allow us to express the metrics as $ \tilde{f}(r)=1+\delta_{\tilde{f}}(r) $ and $ h(r)=1+\delta_{h}(r) $ (where $ \delta_{\tilde{f}}(r) \ll 1 $ and $ \delta_{h}(r) \ll 1 $)  \cite{babichev2016relativistic}.  The physical metric $f$ is calculated from the relation:
\begin{equation}
f(r)=\frac{\tilde{f}(r)}{\tilde{f}_{\infty}}\, ,
\end{equation}
where $\tilde{f}_{\infty}=1/f_c$ is obtained from the numerical integration of the field equations. 

Following the same line of reasoning as above, we perturb the scalar field as $ \phi(r)=\phi_{\infty}+\delta_{\phi}(r) $,  so the scalar field equation (\ref{eq_tov3}) becomes: 
\begin{equation}
	\delta''_{\phi}+\frac{2}{r}\delta'_{\phi}-V''(\phi_{\infty})\delta_{\phi}=0 \, ,
\end{equation}
whose solution has the particularly simple form:
\begin{equation}\label{eq_asym_phi}
	\delta_{\phi}(r)=A_{1}\frac{e^{-r\sqrt{ V''(\phi_{\infty})}}}{r}+A_{2}\frac{e^{r\sqrt{ V''(\phi_{\infty})}}}{r\sqrt{ V''(\phi_{\infty})}}\, .
\end{equation}
This solution involves the mass term  $ \sqrt{V''(\phi_{\infty})} $ where $ A_1 $ and $ A_{2} $ are constants of integration. The regularity of the  scalar field at infinity implies that $ A_{2}=0 $, leaving only the Yukawa suppression term in (\ref{eq_asym_phi}): 
\begin{equation}\label{eq_scalar}
	\phi(r)=\phi_{\infty}+k_{1}\frac{e^{-r\sqrt{ V''(\phi_{\infty})}}}{r} \, .
\end{equation}

\section{Slowly rotating neutron stars}\label{sec_rotation}

In this section we consider slowly rotating NSs. We extend the metric (\ref{eq_static_metric}) to a stationary axially symmetric spacetime \cite{hartle1967slowly}
\begin{equation}\label{eq_rotating_metric}
	ds^{2}=-f(r)dt^{2}+h(r)dr^{2}+r^{2}(d\theta ^{2}+\sin^{2}\theta d {\phi}
	^{2})-2\omega (r,\theta )r^{2}\sin^{2}\theta dtd {\phi} +\mathcal{O}(\Omega
	^{2}) \, ,
\end{equation}
where $\omega (r,\theta )$ accounts for frame dragging effect.

The regular tetrad in the Weitzenb\oo ck gauge that corresponds to the above metric (\ref{eq_rotating_metric}) can be determined by satisfying both conditions (\ref{eq_tetrad_constraints}) and (\ref{eq_tetrad_coef1}) providing that 
 $C^{0}{}_{3}=r^{2} f(r)^{-1/2}\sin^{2}(\theta)\, \omega (r,\theta )$:
\begin{equation}\label{eq_tetrad_rot}
	e^{a}{}_{\mu}=\left( 
	\begin{array}{cccc}
		\sqrt{f(r)} & 0 & 0 & \frac{r^{2}\sin ^{2}\left( \theta \right) }{\sqrt{f(r)}
		}\omega (r,\theta ) \\ 
		0 & \sqrt{h(r)}\sin (\theta )\cos (\phi ) & r\cos (\theta )\cos (\phi ) & 
		-r\sin (\theta )\sin (\phi ) \\ 
		0 & \sqrt{h(r)}\sin (\theta )\sin (\phi ) & r\cos (\theta )\sin (\phi ) & 
		r\sin (\theta )\cos (\phi ) \\ 
		0 & \sqrt{h(r)}\cos (\theta ) & -r\sin (\theta ) & 0
	\end{array}
	\right) \, .
\end{equation}
Since this tetrad inherently satisfies the symmetries discussed in subsection \ref{subsec_axial}, it automatically resolves the antisymmetric field equations \eqref{eq_spiFEQ}, thereby leaving us solely with the symmetric tetrad field equations \eqref{eq_tetFEQ}.

The matter of the rotating star is modeled as a perfect fluid rotating uniformly with angular velocity $\Omega$, defined by:
\begin{equation}
u^{t}=\left[ -\left(g_{tt}+2\Omega g_{t\phi}+\Omega^2 g_{\phi\phi}\right)
\right] ^{-1/2}, \, \quad u^{r}=u^{\theta }=0, \, \quad u^{{\phi} }=\Omega, u^{t}.
\end{equation}
Here, $\Omega$ denotes the angular velocity of the star as observed by an observer at rest at infinity. The NS is assumed to rotate uniformly and slowly enough so as not to significantly perturb other geometrical and physical properties of the star, such as the gravitational field, density, and pressure. This constraint is encapsulated by the condition $\Omega r_{s} \ll c$.

The $(t , {\phi})$ component of the stress-energy tensor of the fluid is given by:
\begin{equation}
	\Theta _{{\phi} }{}^{t}=\frac{r^{2}\sin^{2}\theta }{f}(\Omega -\omega )(\rho
	+P)+\mathcal{O}(\Omega ^{2}) \, .
\end{equation}
We then substitute the tetrad frame \eqref{eq_tetrad_rot} into the field equations \eqref{eq_tetFEQ}, and define the function $\omega=\Omega(1-\bar{\omega})$, where $\bar{\omega}/\Omega$ represents the coordinate angular velocity of the fluid observed by a freely falling observer \cite{hartle1967slowly}. We further expand $\bar{\omega}$ on the basis of the Legendre polynomial $P_{l}$ as follows:
\begin{equation}
\bar{\omega}(r,\theta )=\sum_{l=1}^{\infty }\bar{\omega}_{l}(r)\left( -\frac{
1}{\sin \theta }\frac{dP_{l}}{d\theta }\right).
\end{equation}
This expansion yields the following differential equation for $\bar{\omega}$ in the general case:
\begin{eqnarray}\label{eq_radial}
	\bar{\omega}''_l &+& \left[\frac{4}{r}-\frac{1}{2}\left(\frac{\tilde{f}'}{\tilde{f}}+\frac{h'}{h}\right)-2\frac{F'_{T}}{F_{T}}\right]\bar{\omega}'_l \nonumber \\
&-&
\left[\frac{h\left[\left(l\left(l+1\right)-2\right)F_{T}+r^{2}\left(\rho+P\right)\right]-4 r \sqrt{h}F'_{T}}{r^{2}F_{T}}\right]\bar{\omega}_l-\frac{4  \sqrt{h} F'_{T}}{r F_{T}}=0 \, .
\end{eqnarray}
Assuming the regularity of $F$ and its derivatives for $r=0$, this equation reads for $r\rightarrow 0$ as:
\begin{equation}
		\bar{\omega}_{l}(r) \ra \alpha_{+}r^{S_{+}}+\alpha_{-}r^{S_{-}} \, ,
\end{equation}
where $\alpha_{\pm}$  are constants of integration and
\begin{equation}
S_{\pm }=-\frac{3}{2}\pm 
\sqrt{\frac{9}{4}+ (l(l+1)-2)}.
\end{equation}
 The geometrical regularity of $\bar{\omega}$ at the origin yields $ \alpha_{-}=0 $. Using the density \eqref{eq_model} and substituting Eqs. (\ref{eq_tov1}) and (\ref{eq_tov2}) into \eqref{eq_radial}, we obtain:
\begin{eqnarray}\label{eq_tov5}
	\bar{\omega}''_{l}&+&\left[\frac{4}{r}-\frac{rh\left(\rho+P\right)+r\phi'^{2}-4\xi \sqrt{h} \phi \phi'}{2\left(\kappa+\xi \phi^{2}\right)}
	\right]\bar{\omega}'_l \nonumber \\
	&-& \left[\frac{h(l(l+1)-2)}{r^2}-\frac{8\xi \sqrt{h}\phi \phi'-2rh(\rho+P)}{r\left(\kappa+\xi \phi^2\right)}\right]\bar{\omega}_l -\frac{8\xi  \sqrt{h}\phi \phi'}{r\left(\kappa+\xi \phi^{2}\right)} =0\, .
\end{eqnarray}
Finally, we show that  $ \bar{\omega}(r) $ near the center is given by:
\begin{equation}\label{eq_initial_omega}
	\bar{\omega} = \bar{\omega}_{c}+\frac{6\,\bar{\omega}_{c}\left(\rho_{c}+P_{c}\right)+2\xi \left(\Omega-\bar{\omega}_{c}\right)\phi_{c} V'(\phi_{c})}{15 \left(\kappa +\xi \phi_{c}^{2}\right)}r^{2} +\mathcal{O}(r^4)\, ,
\end{equation}
and that at large $r$ we obtain the solution:
\begin{equation}
\bar{\omega}_{l}(r) \sim \textrm{const.}r^{-l-2}+\textrm{const.}r^{l-1} \, .
\end{equation}
For the space to be flat at large distances the only relevant terms are for $l=1$, and thus a regular solution outside the NS is:
\begin{equation}
\bar{\omega}=1-2I/r^{3},
\end{equation}
where $I$ is the moment of inertia.

Using Eq.(\ref{eq_tov5}) for $ l=1 $ one can find the integral representation of the moment of inertia as:
\begin{equation}\label{eq_inertia}
	I=\frac{1}{3}\int_{0}^{r_{s}} \frac{r^{4}}{\kappa+\xi \phi^2}\sqrt{\frac{h}{\tilde{f}}}\left(\rho+P\right)\bar{\omega} \, dr +\frac{2}{3}\int_{0}^{r_{s}} \frac{r^{3}}{\sqrt{h f}} \frac{\xi \phi \phi'}{\kappa+\xi \phi^2}\left(2\sqrt{h}(1-\bar{\omega})-r\bar{\omega}'\right)dr\, .
\end{equation}
In the numerical integration below, we employ a double shooting method to determine the moment of inertia of the star, accounting for the boundary condition at the star radius:
\begin{equation}
	\bar{\omega}(r_{s})= \left(1-\frac{2I}{r_{s}^{3}}\right), \,\, 
	\bar{\omega}'(r_s)=\frac{6 I}{r_{s}^{4}} \, .
\end{equation}

\section{Numerical results and discussion}\label{sec_profiles}
In this paper, we study the impact of the non-minimal coupling and a specific scalar potential  on the structure of NS in TST theory. The explicit form of the potential  encompasses two positive constants, denoted  $ \mu $ and $ \lambda $ and is given by:
\begin{equation}\label{eq_symmetron}
V(\phi )=-\frac{1}{2}\mu ^{2}\phi ^{2}+\frac{\lambda }{4}\phi ^{4}.
\end{equation}
This potential was primarily used in the study of NSs in the symmetron model \cite{de2021neutron}.

The effective potential in the model under study becomes:
\begin{equation}
V_{\text{eff}}(\phi)=-\frac{1}{2}\mu_{\text{eff}}^2 \phi^2+\frac{\lambda}{4}\phi^4 , 
\end{equation}
where $ \mu_{\text{eff}}^2=\mu^2-\xi T $. Outside the star, at large distances, the torsion scalar vanishes, and the effective potential exhibits critical points, $ \phi_{\infty}=0,\pm \mu / \sqrt{\lambda} $. The first point is unstable, while the others are stable, breaking the $\mathbb{Z}_2$ symmetry. Consequently, the field tends to settle at one of the nontrivial minima of the effective potential. In \cite{arapouglu2019neutron,arapouglu2020neutron}, the choice of considering stable or unstable critical points depends on the asymptotic condition $V(\phi_{\infty})=0$. However, as mentioned earlier, we will consider below the non-trivial stable point $ \phi_{\infty}= \mu / \sqrt{\lambda} $ that satisfies \eqref{eq_box} outside the star with $V(\phi_{\infty})\neq 0$, resulting in an Anti-de Sitter (AdS) spacetime outside the NS. This practice is common in scalar-tensor theories to have $V (\phi_{\infty})\neq 0$, which introduces an effective cosmological constant \cite{sotiriou2012}.

Utilizing the scalar potential (\ref{eq_symmetron}), the initial conditions (\ref
{eq_initial1})-(\ref{eq_adm_general}) and (\ref{eq_initial_omega}) become:
\begin{eqnarray}\label{eq_initial_symmetron1}
	\tilde{f}(r) &=&1+\frac{\left( 2\rho _{c}+6P_{c}+2\mu ^{2}\phi_{c}^{2}-\lambda \phi _{c}^{4}\right) }{12\left( \kappa +\xi \phi
		_{c}^{2}\right) } r^{2} \, ,\\
	h(r) &=&1+\frac{\left( 4\rho _{c}-2\mu ^{2}\phi _{c}^{2}+\lambda
		\phi _{c}^{4}\right) }{12\left( \kappa +\xi \phi _{c}^{2}\right) }r^{2}\, , \label{eq_initial_symmetron2} \\
	\phi (r) &=&\phi _{c}-\frac{1}{6}\left( \mu ^{2}\phi _{c}-\lambda \phi
	_{c}^{3}\right)r^{2} \, , \label{eq_initial_symmetron3} \\
	P(r) &=&P_{c}-\frac{\left( \rho _{c}+P_{c}\right) \left( 2\rho
		_{c}+6P_{c}+2\mu ^{2}\phi _{c}^{2}-\lambda \phi _{c}^{4}\right) }{24\left(
		\kappa +\xi \phi _{c}^{2}\right) } r^{2} \, , \label{eq_initial_symmetron4}\\
		m(r)&=&\frac{\pi \kappa \left( 4\rho _{c}-2\mu ^{2}\phi
		_{c}^{2}+\lambda \phi _{c}^{4}\right) }{2\left( \kappa +\xi \phi
		_{c}^{2}\right) } r^{3} \, ,
\label{eq_initial_symmetron5} \\
	\bar{\omega} &=& \bar{\omega}_{c}+\frac{6\,\bar{\omega}_{c}\left(\rho_{c}+P_{c}\right)+2\xi \left(\Omega-\bar{\omega}_{c}\right)\left(-\mu^{2}+\lambda \phi_{c}^{2}\right)\phi_{c}^{2}}{15 \left(\kappa +\xi \phi_{c}^{2}\right)}r^{2}\, . \label{eq_inital_symmetron5}
\end{eqnarray} 
For completeness, a detailed analysis of the validity of Schwarzschild solutions for an incompressible fluid within the framework of our model is provided in appendix \ref{rho_const}.

The system of equations \eqref{eq_tov1}-\eqref{eq_tov2} and \eqref{eq_tov3}-\eqref{eq_mass}, with the specified initial conditions, is closed by establishing a relationship between the pressure and the energy density of the matter distribution via an equation of state (EOS) that appropriately fits observational data.

The outer crust and inner core of a NS can be characterized by a unified equation of state (EOS) where the pressure and energy density are parametrized by two logarithmic quantities: $\zeta \equiv \log_{10}(P/\text{dyn.cm}^{-2})$ and $\xi \equiv \log_{10}(\rho/\text{g.cm}^{-3})$. Although it is acknowledged that the equations of state employed in NS physics and other astrophysical phenomena are inherently dependent on GR \cite{PhysRevLett.128.202701,2023constraining,wojnar2023fermi}, it is commonly assumed in the literature that using such equations of state in Extended Theories of Gravity does not significantly impact NSs. We then employ a set of four known realistic EOS: SLy, BSk19, BSk20, and BSk21, with their analytical representations provided in \cite{haensel2004analytical,potekhin2013analytical}:
\begin{eqnarray}\label{EOS}
\zeta &=&\frac{a_1+a_2\xi+a_3\xi^2}{1+a_4\xi}d\left(a_5\left(\xi-a_6\right)\right)+\left(a_7+a_8\xi\right)d\left(a_9\left(a_{10}-\xi\right)\right)+\left(a_{11}+a_{12}\xi\right)d\left(a_{13}\left(a_{14}-\xi\right)\right)\nonumber\\
&+&\left(a_{15}+a_{16}\xi\right)d\left(a_{17}\left(a_{18}-\xi\right)\right)
+\frac{a_{19}}{a_{20}^2\left(a_{21}-\xi\right)^2}+\frac{a_{22}}{a_{23}^2\left(a_{24}-\xi\right)^2},
\end{eqnarray}
where $d(x)=\left(e^{x}+1\right)^{-1}$, and the constants $a_i$  with $a_{19},\dots,a_{24}=0$ for SLy  are tabulated in \cite{haensel2004analytical}, and the $a_i$ for the BSk19, BSk20 and BSk21 are given in  \cite{potekhin2013analytical,kase2020neutron}. The analytical representation (\ref{EOS}) fits the tabulated EOS at a precision of order $1-2\%$ in the range $10^{6}< \rho \lesssim2 \times 10^{16}{g\, cm^{-3}}$. The SLy and the BSk family are  considered to be relatively stiff.

For numerical efficiency, it is beneficial to operate with squared roots of the metric functions: $ \tilde{f}(r) \ra \tilde{f}(r)^2,\, h(r) \ra h(r)^2 $. Additionally, instead of the radial coordinate $ r $, we employ the dimensionless parameter $ s $, as defined below \cite{kase2019neutron}:
\begin{equation}
s=\text{ln}\left(\frac{r}{r_{0}}\right),\,
r_{0}=\left(\frac{8 \pi \kappa}{\rho_{0}}\right)^{1/2}=89.664\, \text{km},\, \rho_{0}=m_{n}n_{0}=1.6749 \times 10^{14}\,\text{g/cm}^{-3},
\end{equation}
where $m_n$ denotes the neutron mass and $n_0$ represents the typical number density of the star. Moreover, we introduce the subsequent dimensionless variables for pressure, density and mass:
\begin{equation}\label{eq_rescale1}
P_{m}(r) \ra c^{2}\rho_{0}P(s), \,
\rho_{m}(r) \ra \rho_{0}\rho(s),\, m(r) \ra \frac{4 \pi r_{0}^{3} \rho_0}{3}  m(s).
\end{equation}
Additionally, we rescale the scalar field and the parameters $ \mu $ and $ \lambda $ as follows:
\begin{equation}\label{eq_rescale2}
{\phi} = \tilde{\phi} \sqrt{\kappa},\,
{\mu} = \tilde{\mu}\sqrt{\frac{\rho_{0}}{\kappa}} , \quad {\lambda} = \tilde{\lambda} \frac{\rho_{0}}{\kappa^2 } .
\end{equation}
The form of the modified TOV equations and the initial conditions in terms of the variable $s$ are provided in Appendix \ref{adim_eqs}. Integration of the system of equations is performed in both the interior and exterior regions of the star. In the interior region, we initiate integration at approximately $s_0=-10$, small enough to be considered as the star's center $r=0$, and integrate outwards to the star's surface at $r_s$, where the pressure vanishes ($P(r_s)=0$). In the exterior region, integration is carried out from the star's surface $r_s$ to infinity ($s=15$), with both pressure and density vanishing.

Throughout the numerical process, we employ a double shooting method to calculate the central value of the scalar field $\tilde{\phi}_c$ by setting the scalar field to its value at infinity $ \tilde{\phi}_{\infty}=\tilde{\mu}/\sqrt{\tilde{\lambda}} $, and match the internal and external solutions by imposing the continuity conditions of the metric functions and the continuity of the scalar field and its derivative. Motivated by the fact that the Parametrized Post-Newtonian (PPN) parameters in teleparallel gravity non minimally coupled to a scalar field are identical to the ones in GR \cite{Li_2014,Chen_2015}, the parameters $\tilde{\mu}$ and $\tilde{\lambda}$ are subject to the constraints imposed by solar system tests on the screening mechanism in the symmetron model \cite{hinterbichler2010screening,de2021neutron}. Consequently, the numerical integration is performed using $\tilde{\mu}=3 \times 10^{-8}$ and $\tilde{\lambda}= 10^{-15}$, and we also assume positive values for $\xi$.
\begin{figure}[t]
	\centering 
	\begin{minipage}[tb]{0.4\textwidth}
		\includegraphics[width=\textwidth]{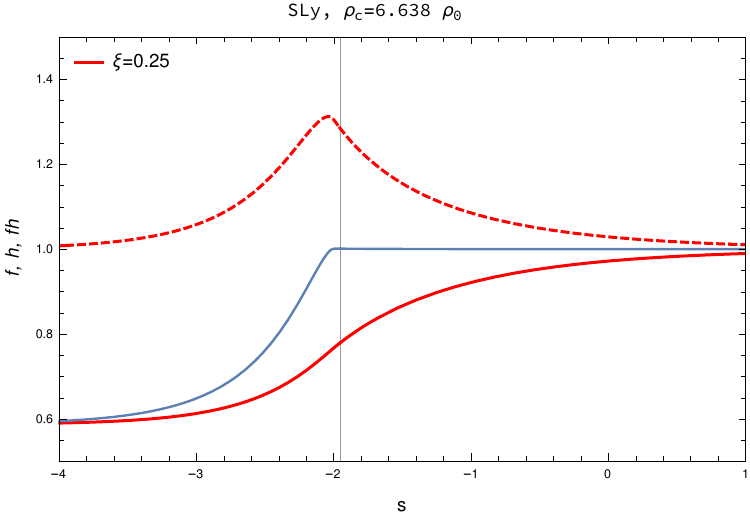}
	\end{minipage}
	\begin{minipage}[tb]{0.4\textwidth}
		\includegraphics[width=\textwidth]{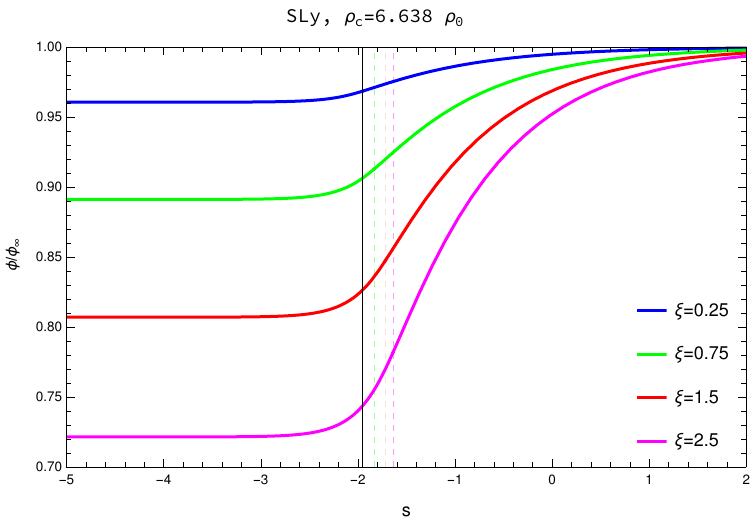}
	\end{minipage}
	\caption{Left: Metric potentials ${f}$ (red solid lines), $h$ (red dashed lines), and their product ${f} h$ (blue solid line) as functions of the radial coordinate $s$ for the SLy EOS and central density $\rho_c=6.638\rho_0$. Unlike the Schwarzschildian exterior solution, this product deviates from unity inside the star. Right: Radial profile of the scalar field rescaled by its asymptotic value ${\Tilde\phi}_{\infty}$  for various values of the coupling parameter $\xi$. The vertical lines mark the location of the star radius $r_{s}$.}
	\label{fig_metrics_scalar}
\end{figure}
\subsection{Mass-radius diagram} 
The metric potentials $f$, $h$, and their product $f\,h$, as well as the scalar field $\tilde{\phi}/\tilde{\phi}_{\infty}=\phi/\phi_{\infty}$, are depicted in Fig. \ref{fig_metrics_scalar} for a central density $\rho_c=6.638,\rho_0$ and $\phi_c=0.685$, using the SLy EOS for various values of the coupling parameter $\xi$. From the profiles, it is evident that the spacetime outside the star is Minkowskian, confirming the validity of the assumptions of a small cosmological constant and weak field approximations used previously. Furthermore, asymptotic flatness can be achieved, even without assuming $V(\phi_{\infty})=0$ a priori.
In the exterior region, the scalar field is described by \eqref{eq_scalar}, showing that the exponential decay term is regulated by a mass term and modulated by the inverse radial distance factor $1/r$. Consequently, the scalar field approaches its asymptotic value $\phi_{\infty}$, or equivalently $\phi/\phi_{\infty} \rightarrow 1$, as $r$ tends to infinity.
\begin{figure}[t]
	\centering 
		\begin{subfigure}[b]{0.4\textwidth}
		\centering
		\includegraphics[width=\textwidth]{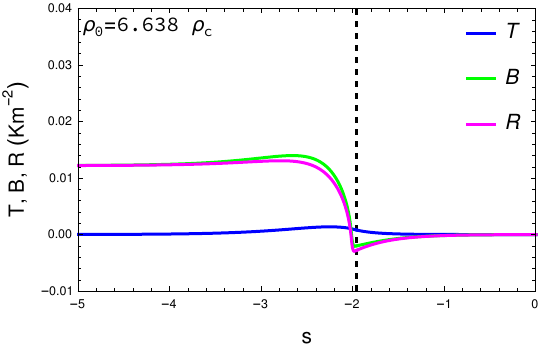}
	\end{subfigure}
	\begin{subfigure}[b]{0.4\textwidth}
		\centering
		\includegraphics[width=\textwidth]{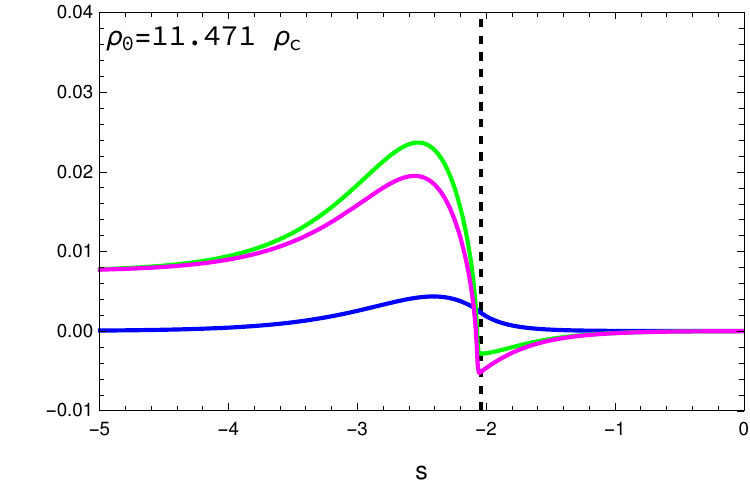}
	\end{subfigure}
	\begin{subfigure}[b]{0.4\textwidth}
		\centering
		\includegraphics[width=\textwidth]{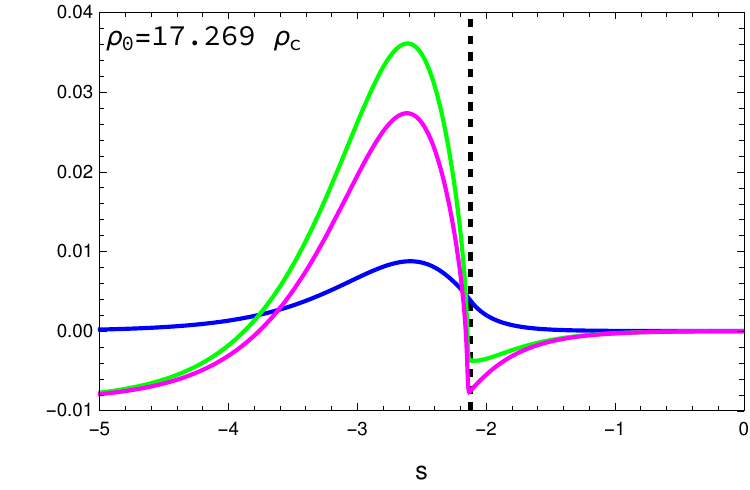}
	\end{subfigure}
	\begin{subfigure}[b]{0.4\textwidth}
		\centering
		\includegraphics[width=\textwidth]{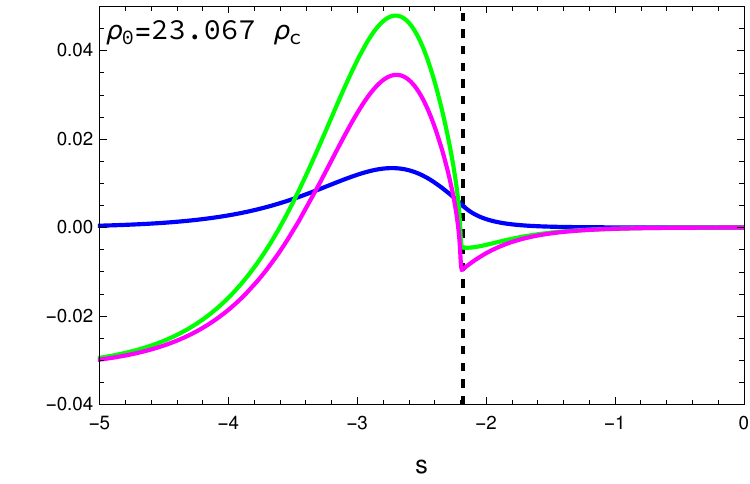}
	\end{subfigure}
		\caption{Scalar torsion $T$, boundary term $B$, and Ricci scalar $R$ as functions of $s$ employing the SLy EOS, for four various values of the central densities $\rho_c$ and coupling parameter $\xi=0.25$.}
	\label{fig_tor_boundary}
\end{figure}
\begin{figure}[t]
   	\centering
   	\includegraphics[width=8cm, height=5cm]{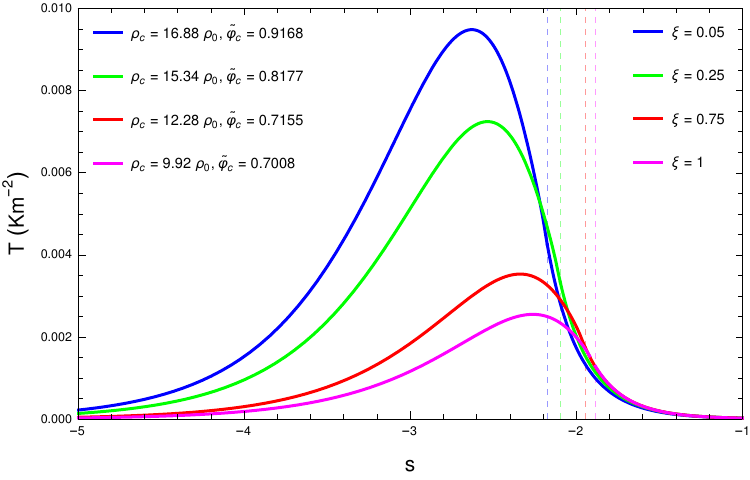}
   		  	\caption{Left: Scalar of torsion vs $s$  employing the SLy EOS, for various values of the central density, the initial scalar field and the coupling parameter. The vertical lines mark the location of the star's radius. }
   	\label{fig_torsion-explanation}
   \end{figure}  
Unlike GR, the dynamic of TEGR is characterized by the torsion scalar $T$, which differs from the Ricci scalar $R$ by the boundary term $B$. In Fig. \ref{fig_tor_boundary}, we illustrate the behaviors of $T$, $B$, and $R$ with respect to the radial coordinate. 
It is evident that $T(r=0)=0$, while $B$ starts from a positive value for lower central densities and becomes increasingly negative for higher central densities. We note that within the stellar mass, both $T$ and $B$ increase until reaching maximum values near the star's surface, after which they begin to decrease until reaching the stellar surface. Notably, the increase in $B$ is more pronounced than that of $T$. Consequently, the contribution of the boundary term outweighs that of torsion scalaire in the non-minimally coupled TST model, resulting in significant deviations from GR and the non-minimally coupled scalar field-Ricci scalar model, as we will show below.

In Fig. \ref{fig_torsion-explanation}, we show the radial variation of the torsion scalar for various values of the coupling parameter employing the SLy EOS. We observe that the larger the coupling parameter, the less significant the growth of the torsion scalar is, before reaching a maximum value, and then starting a faster decrease near the surface of the star, eventually asymptotically reaching zero at radial infinity. Importantly, the torsion scalar in the immediate vicinity of the outer surface of the star is not zero due to coupling with the scalar field.

\begin{figure}[t]
   	\centering
   	\begin{minipage}[htb]{0.45\textwidth}
   		\includegraphics[width=\textwidth]{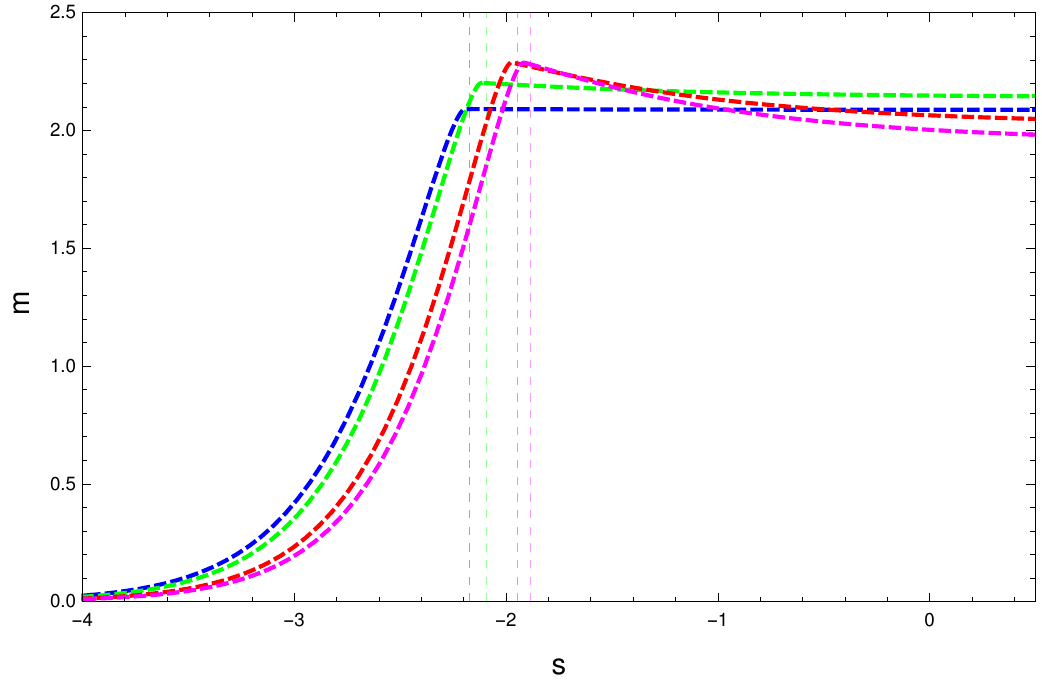}
   	\end{minipage}
   	\begin{minipage}[htb]{0.45\textwidth}
   		\includegraphics[width=\textwidth]{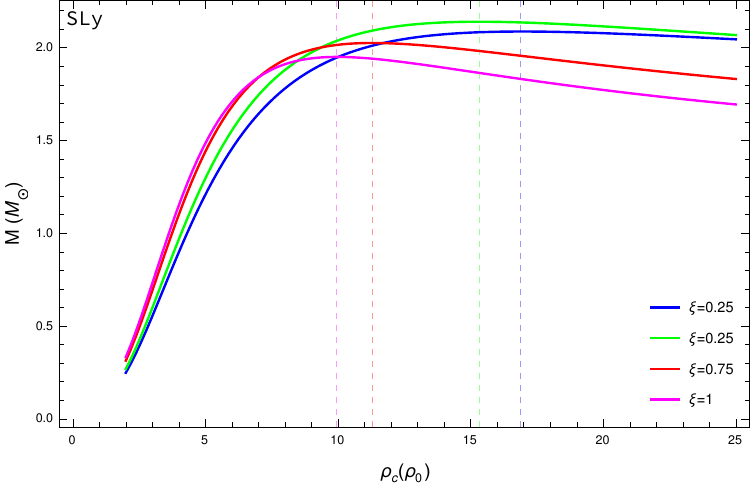}
   		\end{minipage}
   	   	\caption{Left: The mass $m(s)$ vs $s$. Right: The mass $m(s)$ vs the central density $\rho_c$. In both panels, we use the SLy EOS, and the vertical lines mark the location of the star's radius.}
   	\label{fig_Mmax-explanation}
   \end{figure}

In the left panel of Fig. \ref{fig_Mmax-explanation}, we show the variation of the stellar mass $M_s$ with the radial coordinate, for varying values of the scalar coupling parameter $\xi$. We note a characteristic increase in the stellar mass, fueled by the energy density inherent within the star, consistent with GR predictions. However, this increase is moderated by a reduction in the effective gravitational constant and the scalar of torsion, particularly pronounced with higher coupling constants. As a result, the mass function reaches a local maximum near the surface of the star, differing from the asymptotic gravitational value expected from GR. Outside the star, at large distances from its surface, the torsion scalar approaches asymptotically zero, as shown in Fig. \ref{fig_torsion-explanation}. Consequently, the stellar mass diminishes until it matches its gravitational value. This effect arises from the weakening of gravity, which permits the existence of stars with lower mass but larger radii. In the right panel of Fig. \ref{fig_Mmax-explanation}, we show the dependence of stellar mass
on the central density of the star. We observe that with stronger coupling, the central density increases, while the mass decreases. This happens because the coupling between the scalar field and torsion scalar works to reduce the total mass of the star beyond its surface. A similar trend in stellar mass has been noted in Bekenstein’s Tensor-Vector-Scalar (TeVeS) theory \cite{PhysRevD.70.083509, PhysRevD.71.069901, Lasky_2008}. 

\begin{figure}[t]
	\centering
	\begin{subfigure}[b]{0.4\textwidth}
		\centering
		\includegraphics[width=\textwidth]{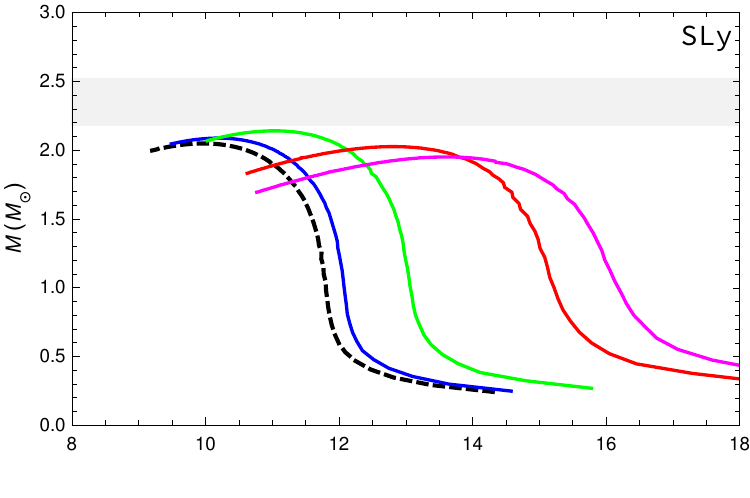}
	\end{subfigure}
	\begin{subfigure}[b]{0.4\textwidth}
		\centering
		\includegraphics[width=\textwidth]{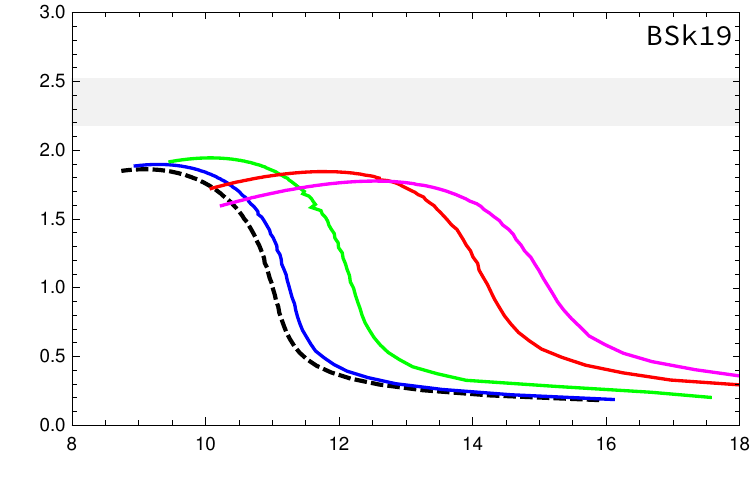}
	\end{subfigure}
	\begin{subfigure}[b]{0.4\textwidth}
		\centering
		\includegraphics[width=\textwidth]{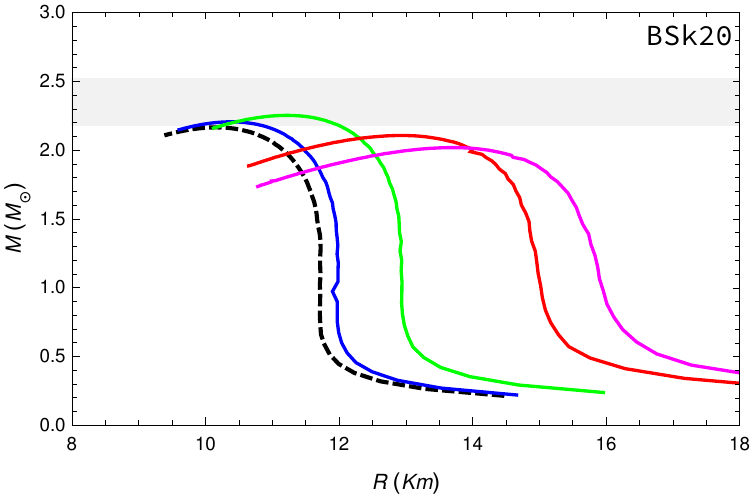}
	\end{subfigure}
	\begin{subfigure}[b]{0.4\textwidth}
		\centering
		\includegraphics[width=\textwidth]{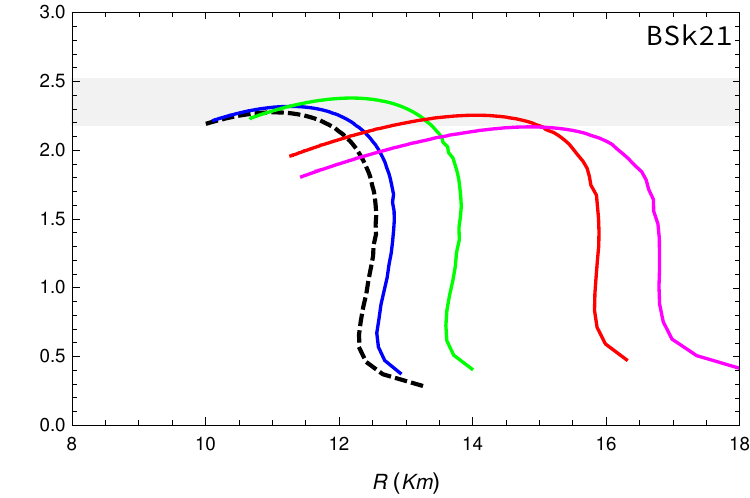}
	\end{subfigure}
	\caption{Mass-radius relation for $\tilde{\mu}=3 \times 10^{-8}$ and $\tilde{\lambda}= 10^{-15}$, corresponding to $\tilde{\phi}_{\infty}=0.9487$, and various values of the coupling parameter $\xi$. Curves are plotted for GR (black dashed line), $\xi=0.05$ (blue line), $\xi=0.25$ (green line), $\xi=0.75$ (red line), and $\xi=1$ (magenta line). The light gray shaded area denotes the measured mass of PSR J0952-0607 \cite{Romani_2022}.}
	\label{fig_mass_radius}
\end{figure}

Understanding how any gravitational theory differs from GR involves examining how the mass of a NS correlates with its radius. We analyze this correlation, considering the set of the four EOSs and various coupling parameters $ \xi $. The results, depicted in Fig. \ref{fig_mass_radius}, are consistent with constraints from the GW170817 event \cite{abbott2017gw170817, Margalit_2017}. We observe that we have less massive NSs with larger radii for values of the scalar coupling $\xi > 0.25$.  Specifically, according to the BSk20 and BSk21 EOSs, the maximum mass is consistent  with the measured mass of PSR J0952-0607 \cite{Romani_2022}. Numerical integration demonstrates that the SLy, BSk20, and BSk21 EOSs can accommodate NSs with masses exceeding $2\,M_{\odot}$. Our findings corroborate the maximum mass limits established in prior studies \cite{Rezzolla_2018}, which range from $2.01^{+ 0.04}_{-0.04}$ to $2.16^{+0.17}_{- 0.15} \, M_{\textrm{TOV}}/M_{\odot}$, derived from observed gravitational events. These limits are based on the quasi-universal relationship between the maximum mass of non-rotating stellar models and the maximum mass supported through uniform rotation, where $M_{\textrm{TOV}}$ represents the maximum mass supported by non-rotating NSs.

We'll now use the universal relations established in \cite{Konstantinou_2022} to estimate how rotation affects the mass of the NS. These relations offer formulas for the fractional increase in mass and radius of a rotating star, based on its rotational frequency and the properties of its non-rotating equivalent. The NS under consideration is PSR J0952–0607 \cite{Romani_2022}, a millisecond pulsar in a binary system, with a mass of $2.35\pm 0.17 M_{\odot}$ and a spin frequency of $707$ Hz. By applying the empirical formulas to the non-rotating NS with $M=2.378 M_{\odot}$ and $R=12.211 \text{ km}$ using the BSk21 EOS, we find that the equivalent rotating NS at $707$ Hz has a mass and radius of $M=2.451 M_{\odot}$ and $R=12.57$ km, respectively. The relative increase in mass and radius due to rotation is calculated to be $3\%$, which falls below the measurement precision. This indicates that the effects of rotation are considerably smaller compared to the precision with which the mass and radius of the PSR J0952–0607 pulsar are measured. 
\begin{figure}[t]
   	\centering
   	   		\includegraphics[width=8cm, height=5cm]{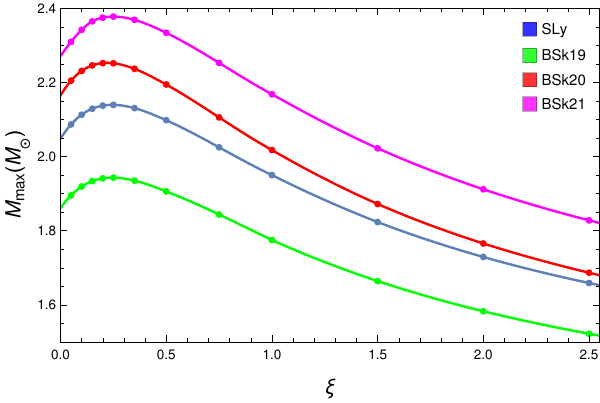}
   	\caption{Evolution of the NS maximum mass with respect to the coupling parameter $\xi$  for the set of the four EOSs.}
   	\label{fig_Mmax}
   \end{figure}
   
Intriguingly, Fig. \ref{fig_Mmax} reveals a notable trend as we vary the coupling parameter from $\xi=0$ (representing no coupling, akin to the quintessence GR model) to $\xi=2.5$. Beyond a critical value, around $\xi\simeq 0.25$, the maximum mass of NSs begins to decrease, suggesting an optimal range where the maximum NS mass is maximized. This departure from GR is highlighted by a distinct feature: the stellar mass $M_s$, contained within the star's radius, exceeds the gravitational mass perceived by an observer at infinity. It's noteworthy that the rate of gravitational mass decrease relative to the stellar mass is estimated to be approximately $2.7\%$ for $\xi\approx0.25$. Additionally, the threshold of $\xi\approx 0.25$ appears to be nearly independent of the EOS.

It is not clear why we have this particular behavior at $\xi\simeq 0.25$, but the physical mechanism can be drawn as follows: the decrease in the effective gravitational constant is less significant inside the star than outside due to the evolution of the scalar field, as can be seen in the right panel of Fig. 1, where the increase in the scalar field is more pronounced with increasing values of the coupling parameter.

\begin{figure}[t]
   \centering
   	   		\includegraphics[width=8cm, height=5cm]{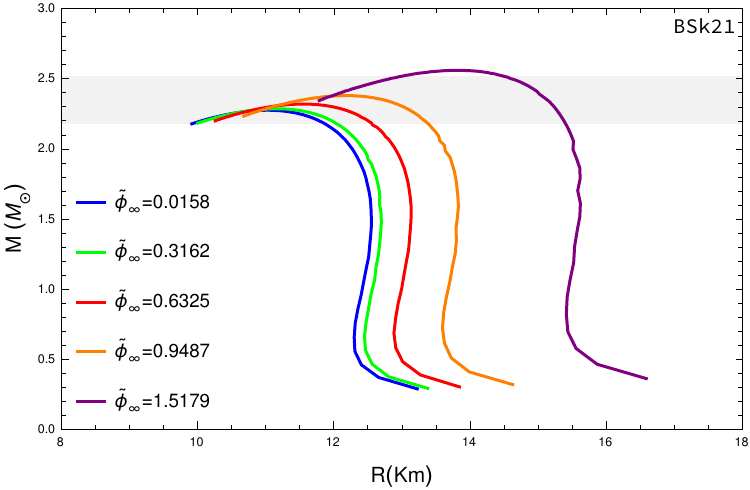}
   	\caption{The mass-radius diagram is depicted for the BSk21 EOS considering various values of the field at infinity $\tilde{\phi}_{\infty}$, and with $\xi=0.25$. The light gray shaded area represents the measured mass of PSR J0952-0607 \cite{Romani_2022}.}
   	\label{fig_Rmin_coupling}
   \end{figure}

In Fig. \ref{fig_Rmin_coupling}, we depict the variations in the mass-radius diagram while altering the value of the scalar field at spatial infinity for the coupling $\xi=0.25$. This exploration was conducted for BSk21 by holding $\tilde{\lambda}=10^{-15}$ and adjusting $\tilde{\mu}$ to $\left( 0.05,\, 1,\,2,\, 3,\,4.8\right)\times 10^{-8}$. Notably, we observe that the larger $\tilde{\mu}$ is, the larger the maximum mass and radius are. Thus, the mass term of the scalar field in massive scalar-torsion theories leaves a larger parameter space unconstrained, resulting in more pronounced deviation from GR.

All the data regarding the maximum mass and their corresponding radii for the set of EOSs are compiled in Table \ref{tab_mass_radii}, where the last two columns present the characteristics of the equivalent NS spinning at $707$ Hz.. In this table, it is observed that for $\xi=0.25$, as the mass of the NS increases, the central density decreases, except for the BSk19 EOS. This EOS is considered as  unreliable, because  it does not allow for NS with mass $M\geq 2M_{\odot}$ within GR \cite{cisterna2016slowly}.  Additionally, the central value of the scalar field remains unaffected by the EOS. Furthermore, the TST model with non-minimal coupling accommodates NSs with smaller central densities compared to GR.

    \begin{table*}[h]
   \centering
\hspace{2cm}
\small
\centerline{\begin{tabular}{c|c|c|c|c|c|c|c|c|c|} 
 \cline{2-10}
&\multicolumn{3}{|c|}{GR}&\multicolumn{4}{|c|}{NRNS ($\xi=0.25$)}&\multicolumn{2}{|c|}{RNS ($\xi=0.25$)}
\\ \hline

\multicolumn{1}{|c|}{EOS} & $\rho_c(\rho_0)$ & $ M_{max}(M_{\odot})$ & $R (Km)$ & $\rho_c(\rho_0)$ & $\phi_c(\kappa^{1/2})$ & $M_{max}$ & $ R $ & $M_{max}$ & $ R $ 
\\ \hline
 
\multicolumn{1}{|c|}{SLy}& 17.076 & 2.048 & 9.97 & 15.27 & 0.838 &  2.14 & 11.037 & 2.194 & 11.3  
\\ \hline
 
 \multicolumn{1}{|c|}{BSk19} & 20.748 & 1.86 & 9.09 & 19.20 & 0.837 & 1.943 &  10.052 & 1.984 & 10.25 
\\ \hline

\multicolumn{1}{|c|}{BSk20} & 16.11 & 2.165 & 10.153 & 14.75 & 0.83 & 2.253 &  11.117 & 2.306 & 11.38 
\\ \hline

\multicolumn{1}{|c|}{BSk21}& 13.6 & 2.274 & 11.037 & 12.44 & 0.84 & 2.378 & 12.211  & 2.451 & 12.57 
\\ \hline

\end{tabular}}

 \caption{Maximum mass and the corresponding radii are provided for GR, the non-rotating NS (NRNS), and the equivalent rotating NS (RNS) spinning at a frequency of $\nu=707$ Hz for $\xi=0.25$. Additionally, we present the central density for GR, and for the scalar-torsion model, we display both the central density and the central value of the scalar field.} \label{tab_mass_radii}
\end{table*}

\begin{figure}[t]
	\centering
	\begin{subfigure}[b]{0.45\textwidth}
		\centering
		\includegraphics[width=\textwidth]{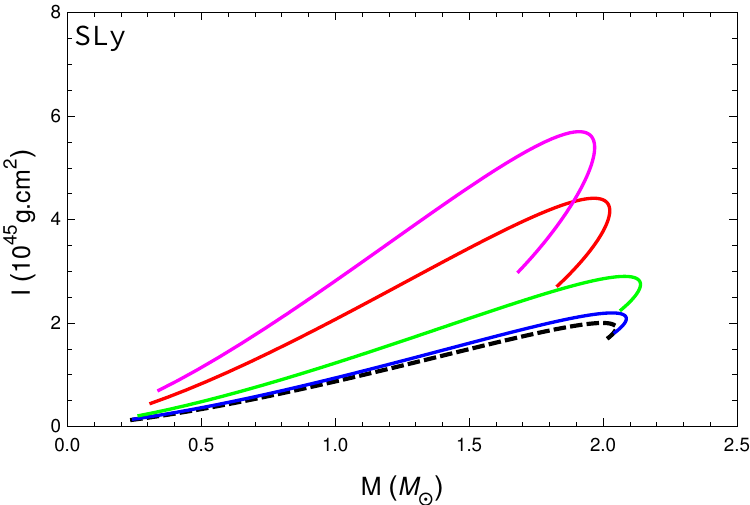}
	\end{subfigure}
	\begin{subfigure}[b]{0.45\textwidth}
		\centering
		\includegraphics[width=\textwidth]{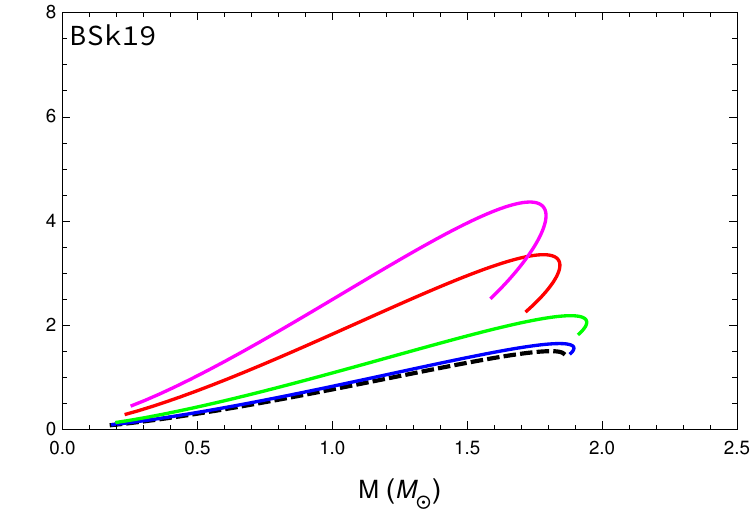}
	\end{subfigure}
	\begin{subfigure}[b]{0.45\textwidth}
		\centering
		\includegraphics[width=\textwidth]{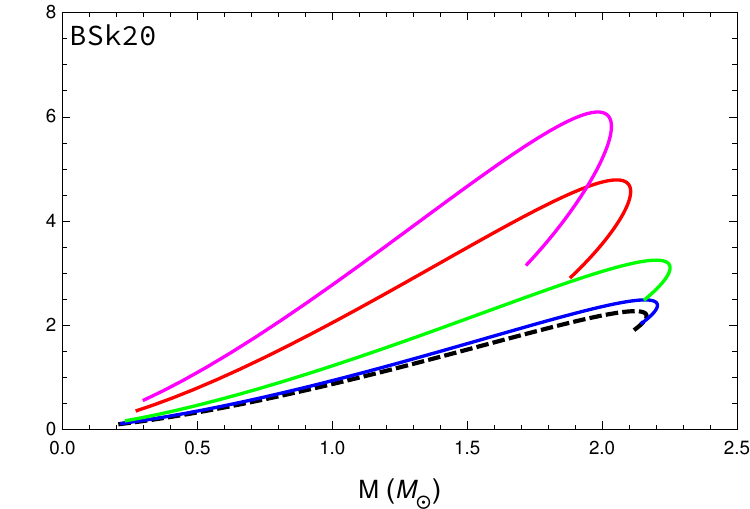}
	\end{subfigure}
	\begin{subfigure}[b]{0.45\textwidth}
		\centering
		\includegraphics[width=\textwidth]{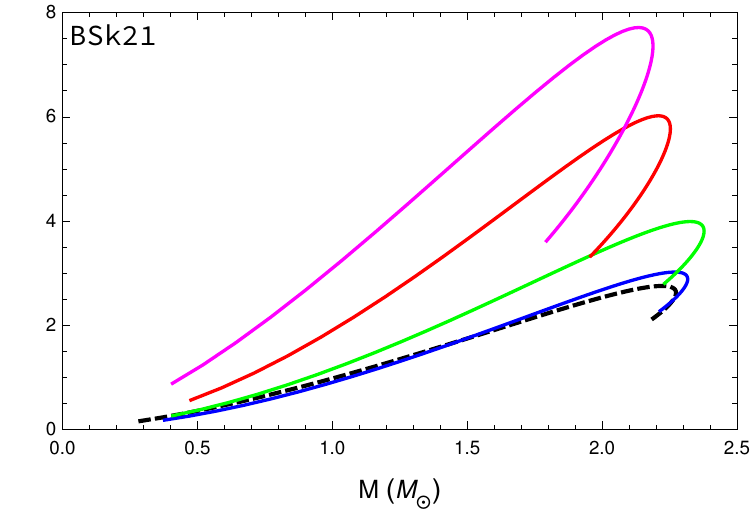}
	\end{subfigure}
	\caption{The moment of inertia-mass relations are plotted for the set of four EOSs and the same parameters as in Fig. \ref{fig_mass_radius}. The curves for GR are represented by black dashed lines, while $\xi=0.05$ is depicted in blue, $\xi=0.25$ in green, $\xi=0.75$ in red, and $\xi=1$ in magenta.}
	\label{fig_inertia_mass}
\end{figure}

In Fig. \ref{fig_inertia_mass}, we depict the variation of the moment of inertia as a function of the mass. Due to its quadratic dependence on radius, the moment of inertia shows a nearly linear increase with the mass of NS, particularly for smaller masses, while it declines rapidly once the maximum mass is approached \cite{breu2016maximum}. Interestingly, we note that unlike the maximum mass of the NS, the moment of inertia of the NS attains larger values as the non-minimal coupling parameter increases, even surpassing the threshold of $\xi\approx0.25$. This implies that the presence of scalar fields affects the moment of inertia's behavior at greater distances, consequently enhancing the rotation speed of the star compared to NS in GR.
\subsection{Surface redshift function}
\begin{figure}[t]
   	\centering
   	\begin{minipage}[htb]{0.45\textwidth}
   		\includegraphics[width=\textwidth]{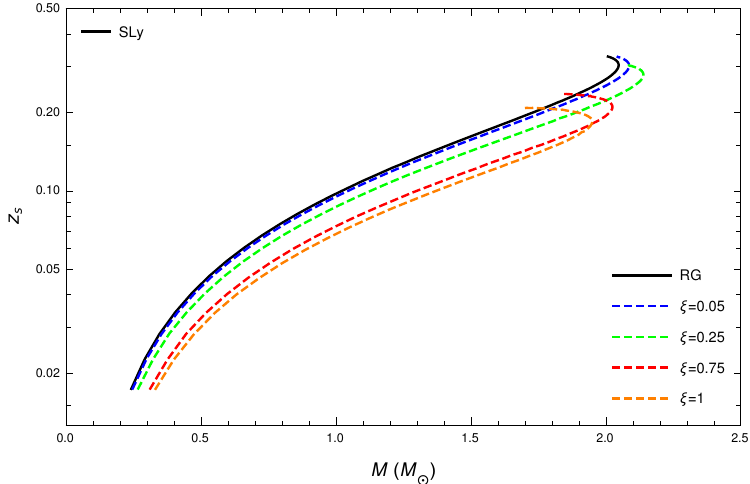}
   		\end{minipage}
   	\begin{minipage}[htb]{0.45\textwidth}
   		\includegraphics[width=\textwidth]{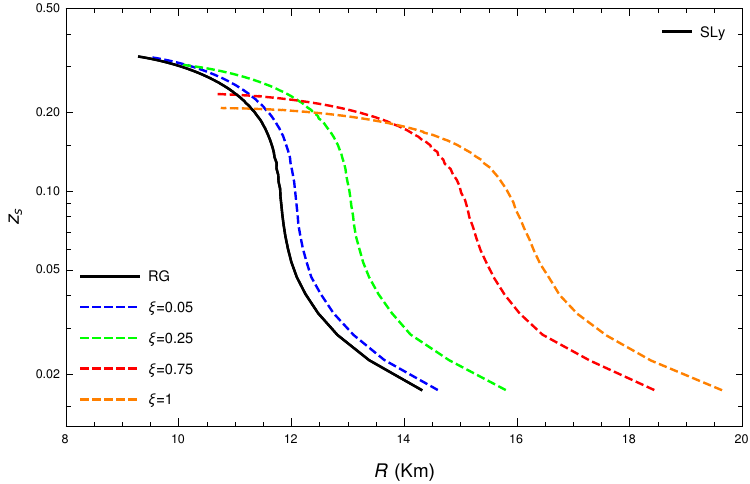}
   	\end{minipage}
   	\caption{Left: Redshift as function of the gravitational mass mass for   various values  of $\xi$ employing the SLy EOS. Right: Redshift as function of the star radius.}
   	\label{fig_redshift}
   \end{figure}

Here, we explore the surface of gravitational redshift, which is an observable quantity revealing the  interaction between gravity, light, and the structure of NSs. It is defined by:
\begin{equation}
z_s = \sqrt{\frac{g_{tt}(\infty)}{g_{tt}(r_s)}} - 1
\end{equation}
Here, $z_s$ is calculated using the gravitational mass $M$. The deviation $\Delta z_s = z_{s} - z_{s,\,gr}$ serves as a tool for distinguishing modified theories of gravity from GR  \cite{DeDeo_2003}. In Fig. \ref{fig_redshift}, we illustrate the redshift at radial infinity of a photon emitted from the NS surface as a function of the ADM mass and the star radius for various values of $\xi$, employing the SLy EOS. We observe that the degeneracy in $z_s$ resulting from mass impedes differentiation between GR and TST theories. However, this degeneracy can be resolved by considering the predicted radius for observational allowed NS masses. For small values of the coupling parameter, $\xi \lesssim 0.05$, deviations from GR are minor. However, for higher values, it is notable that as the coupling parameter increases, the redshift decreases. Thus, for $\xi > 0.25$, we observe redshift $z_s < 0.3$. It is noteworthy that for photons emitted at the redshift $z_s \gtrsim 0.24$, NSss for $\xi \gtrsim 0.75$ are excluded.
\subsection{Universality relations}
The universality relations governing the moment of inertia, mass, and radius of NS were initially introduced in  \cite{Ravenhall1994} and extensively explored in  \cite{breu2016maximum}. These relations offer two distinct normalizations for the moment of inertia: $\widetilde{I}=I/(M r_{s}^2)$ and $\overline{I}=I/M^3$, both expressed in terms of the star's compactness, ${\cal{C}}=M/r_{s}$, as follows:
\begin{eqnarray}
\widetilde{I}&=&a_{0}+a_{1}{\cal{C}}+a_{4}{\cal{C}}^4 , \label{eq_normalized}\\
\overline{I}&=&b_{1}{\cal{C}}^{-1}+b_{2}{\cal{C}}^{-2}+b_{3}{\cal{C}}^{-3}+b_{4}{\cal{C}}^{-4} , \label{eq_dimensionless}
\end{eqnarray}
where the coefficients $a_{i}$ and $b_{i}$ are fitted constants provided in Tables (\ref{tab_univ1}) and (\ref{tab_univ2}) for $\xi=0.05$ and $\xi=0.25$. The last column in each table presents the reduced chi-square values, $\chi^{2}_{red}$, comparable to those reported in \cite{staykov2016moment}.

\begin{figure}[H]
	\centering
	\begin{minipage}[htb]{0.48\textwidth}
		\includegraphics[width=\textwidth]{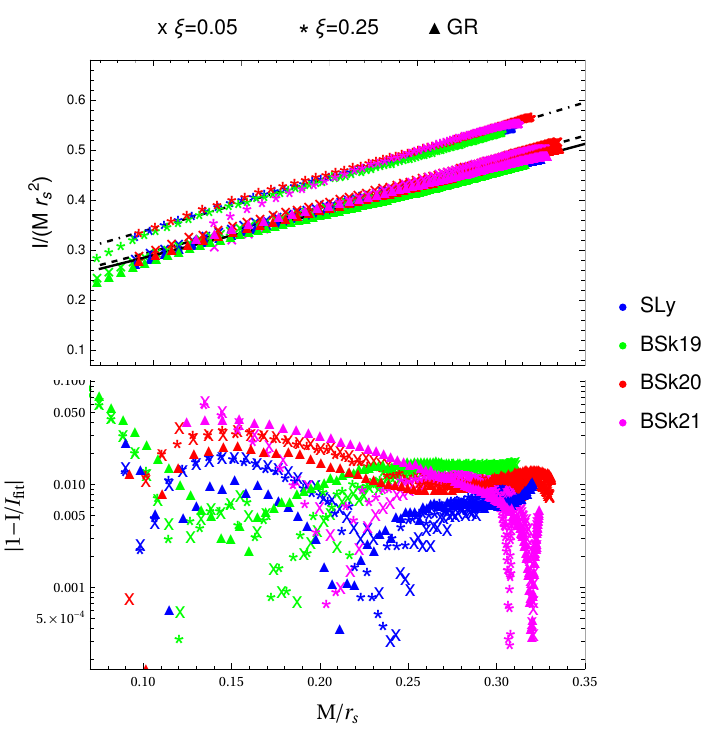}
	\end{minipage}
	\begin{minipage}[htb]{0.48\textwidth}
		\includegraphics[width=\textwidth]{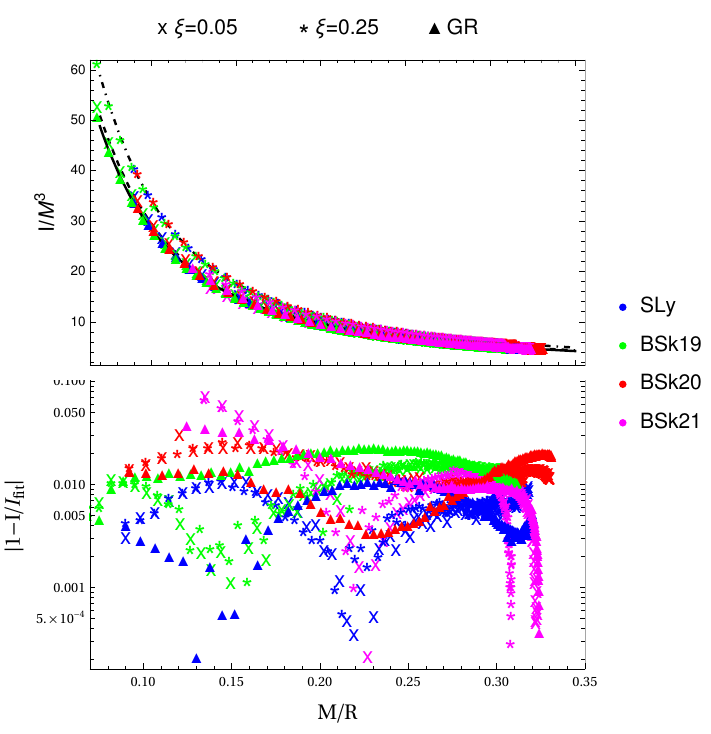}
	\end{minipage}
\caption{The different normalizations of the moment of inertia in terms of the stellar compactness in slow rotation approximation for the model (\ref{eq_model}). The solid, dashed and dot-dashed black lines are the polynomial fits for GR and scalar torsion theory with $\xi=0.05$ and $\xi=0.25$ , respectively. In the lower panels we plot the deviation from universality $ \left| 1-I/I_{\text{fit}} \right| $.}
\label{fig_universal}
\end{figure}

In Fig. \ref{fig_universal}, we depict $\widetilde{I}$ and $\overline{I}$ as functions of compactness for both GR and  TST with non minimal coupling for $\xi=0.05$ and $\xi=0.25$. Notably, both $\widetilde{I}$ and $\overline{I}$ exhibit higher values compared to those predicted by GR. In the left panel, a deviation from GR is observed for $\xi=0.25$, particularly with the BSk21 EOS. Conversely, in the right panel, deviations in the normalized moment of inertia $\overline{I}$ from GR significantly diminish.

Further examination in the lower panels involves measuring the relative deviations $\left| 1-I/I_{\text{fit}} \right|$, revealing errors below $10\%$ for both normalized moments of inertia, dropping below $1\%$ for ${\cal{C}}=0.3$. Within the compactness range of $0.1$ to $0.3$, polynomial fitting suggests an approximate increase of $3\%$ and $15\%$ for $\widetilde{I}$ and $\overline{I}$, respectively, over their GR counterparts. We can conclude that the EOS universality of $\widetilde{I}$ and $\overline{I}$ is upheld in TST with non-minimal coupling, at least as effectively as in GR.

\begin{table}[H]
	\centering
	\small
	\begin{tabular}{|c|c|c|c|c|}
		\hline
		& $ a_{0} $ & $ a_{1} $ & $ a_{4} $ & $ \chi^{2}_{red} $ \\
		\hline
		$ \text{GR} $ & $ 0.200 \pm 0.002 $ & $ 0.884 \pm 0.013 $ & $ 0.401 \pm 0.218 $ & $ 3.13 \times 10^{-5} $ \\
		\hline
		$ \xi=0.05 $ & $ 0.206 \pm 0.002 $ & $ 0.911 \pm 0.013 $ & $ 0.505 \pm 0.219 $ & $ 3.02 \times 10^{-5} $ \\
		\hline
		$ \xi=0.25 $ & $ 0.242 \pm 0.002 $ & $ 1.009 \pm 0.016 $ & $ 0.277 \pm 0.308 $ & $ 3.2 \times 10^{-5} $ \\
		\hline
	\end{tabular}
	\caption{The best fitted coefficients $a_i$ obtained from the data with the relation (\ref{eq_normalized}).}
	\label{tab_univ1}
\end{table}
\begin{table}[H]
	\centering
	\small
	\begin{tabular}{|c|c|c|c|c|c|}
		\hline
		& $ b_{1} $ & $ b_{2} $ & $ b_{3} $ & $ b_{4} $ & $ \chi^{2}_{red} $ \\
		\hline
		$ \text{GR} $ & $ 0.843 \pm 0.017 $ & $ 0.215 \pm 0.008 $ & $ 0.0027 \pm 0.001 $ & $ -0.0003 \pm 0.00004 $ & $ 1.4 \times 10^{-2} $ \\
		\hline
		$ \xi=0.05 $ & $ 0.974 \pm 0.021 $ & $ 0.171 \pm 0.009 $ & $ 0.007 \pm 0.001 $ & $ -0.0004 \pm 0.00005 $ & $ 1.92 \times 10^{-2} $ \\
		\hline
		$ \xi=0.25 $ & $ 1.049 \pm 0.025 $ & $ 0.211 \pm 0.011 $ & $ 0.007 \pm 0.001 $ & $ -0.0005 \pm 0.00005 $ & $ 2.41 \times 10^{-2} $ \\
		\hline
	\end{tabular}
	\caption{The best fitted coefficients $b_i$ obtained from the data with the  relation (\ref{eq_dimensionless}).}
	\label{tab_univ2}
\end{table}

\section{Conclusion}\label{sec_conclusion}
In this paper, we have studied the existence of realistic NSs in the the framework of scalar torsion theories, focusing on models featuring non-minimal coupling between the scalar field and the torsion scalar, along with a scalar potential. Notably, we distinguished our analysis by parameterizing the theory with the coupling parameter $\xi$, as well as the potential parameters $\mu$ and $\lambda$.

Analytically, we demonstrated that the spacetime outside the star exhibits Schwarzschild Anti-de-Sitter type (SAdS), a consequence of the non vanishing potential at spatial infinity. Within the star, we modeled the stellar matter as a perfect fluid, employing various EOS, including SLy, BSk19, BSk20, and BSk21 EOSs. 
Expanding our investigation, we incorporated the slow rotation approximation using the regular Weitzenböck tetrad to generate an axially symmetric spacetime and derive the radial rotation equation. Numerical integration of the resulting partial differential equations allowed us to obtain a mass-radius relation for the star, consistent with the constraints imposed by the GW170817 event. 
A significant finding emerged from our study: as the coupling parameter surpassed a critical threshold,  $\xi\simeq 0.25$, we observed a decrease in the maximum mass of the NS. This observation suggested the presence of an optimal range where the maximum mass is maximized, marking a departure from General Relativity. Specifically, our study reveals that at $\xi=0.25$, the maximum mass of the star peaked at $M = 2.37 M_{\odot}$ for the BSk21 EOS, is consistent with the mass of PSR J0952-0607. 
Furthermore, our investigation revealed a notable deviation from the expected behavior in GR: the stellar mass function exhibited a local maximum near the star's surface, different from the asymptotic gravitational value. This discrepancy was attributed to the predominant influence of the coupling term outside the star in the just vicinity of the star surface, resulting in a gradual decrease in stellar mass until it aligned with its gravitational value. This phenomenon, driven by the weakening of gravity, allows the existence of stars with lower mass but larger radii. It also confirms that the external structure of the spacetime is altered by the scalar field.

Additionally, we observed that as the mass of the NS increased, the central density decreased. Moreover, the central value of the scalar field remained unaffected by the EOS. Furthermore, we found that TST theory with non-minimal coupling accommodated NSs with smaller central densities compared to GR. Moreover, massive scalar-torsion theories left a larger parameter space unconstrained, resulting in more pronounced deviations from GR.

We also observed that NSs display a surface redshift that deviates from GR as the coupling parameter increases. Particularly noteworthy is that NSs in TST theory with non-minimal coupling become undetectable for $\xi \gtrsim 0.75$, for values of the  surface redshift  $z_c \gtrsim 0.24$.
 
Our study also investigated the evolution of the moment of inertia with the mass of the star, revealing that the presence of scalar fields influences the behavior of the moment of inertia at larger distances compared to GR results. Finally, we examined the evolution of the normalized moment of inertia and found good universality relations for both GR and the TST theory with non minimal coupling. The deviations from universality were found to be less than $10\%$.

In summary, our findings provide valuable insights into the intricate interplay between various parameters in TST theories with non minimally coupled scalar field, shedding light on the structural properties of NSs and highlighting departures from conventional gravitational theories like GR. The results we have obtained can be aligned with those already obtained in the cosmological context where teleparallel dark energy with non-minimal coupling agrees perfectly with observational data for $0<\xi<0.2$ for quartic and exponential potentials, and mimics the $\Lambda$CDM model. Indeed, in this scenario, the equation of state parameter of dark energy $w_{DE}$ is approximately constant and has a value $\simeq -0.98$, very close to the cosmological constant \cite{cai2016f}. Our results once again confirmed the structural richness of the TST with non-minimal coupling compared to its counterpart in RG.

\appendix
\section{Interior solutions with an incompressible fluid}\label{rho_const}
In this appendix, we investigate the validity of Schwarzschild interior solutions within the framework of the scalar torsion model, with non-minimal coupling and a scalar potential, for an isotropic star with constant density. Our analysis is grounded on the observation  that the scalar field exists against the background of an incompressible fluid star. Consequently, the field equations derived in Sec.  \ref{sec_static} account for fluctuations around $\phi=0$ up to order ${\cal{O}}\left(\phi^2,\phi\phi',\phi'^2,\phi^4\right)$, where, at the leading order, only the scalar field equation becomes non-trivial. This approximation is validated by examining the profiles of the torsion scalar and the scalar field within the star, which are nearly negligible except in the vicinity of the star's surface, as depicted in Fig. \ref{fig_metrics_scalar}. At the background, the scalar field is null, and the solutions to the Einstein field equations inside the star are governed by the Schwarzschild RG interior solutions:
\begin{eqnarray}
f(r)&=&\left[\frac{3}{2}\left(1-2 {\cal{C}}\right)^{1/2}-\frac{1}{2}\left(1-{\cal{C}}\frac{r^2}{r_s^2}\right)^{1/2}\right]^2,\label{fs}\\
h(r)&=&\left(1-2{\cal{C}}\frac{r^2}{r_s^2}\right)^{-1} \label{hs},\\
P(r)&=& \rho_c c^2\frac{\left(1-2{\cal{C}}\frac{r^2}{r_s^2}\right)^{1/2}-\left(1-2{\cal{C}}\right)^{1/2}}{3\left(1-2{\cal{C}}\right)^{1/2}-\left(1-2{\cal{C}}\frac{r^2}{r_s^2}\right)^{1/2}}\label{ps},
\end{eqnarray}
where ${\cal{C}}=MG/c^2 r_s$ is the compactness of the star. Substituting these solutions in \eqref{eq_torsion_calar} and \eqref{eq_boundary_term}, we show that the scalar of torsion and the boundary term to order  ${\cal{O}}\left(\left({\cal{C}}\frac{r^2}{r_s^2}\right)^3\right)$ are given  by:
\begin{equation}
T\simeq0,\,
B\simeq\frac{7 {\cal{C}}^2 r^2}{r_s^4}-\frac{27 {\cal{C}}^2}{2 r_s^2}+\frac{9 {\cal{C}}}{r_s^2}.
\end{equation}
We observe that $T$ is suppressed compared to $B$, and that these expressions are consistent with the evolution of $T$ and $B$ shown in Fig. \ref{fig_tor_boundary} for low central density. It is clear that the GR interior solutions cannot reproduce the evolution of $T$ and $B$ near the star surface.

Under these assumptions, the scalar field equation is expanded as follows:
\begin{equation}
\phi''+\left(\frac{2}{r}+\frac{\tilde{f'}}{2\tilde{f}}-\frac{h'}{2h}\right)\phi'+\mu^2\phi+{\cal{O}}\left(\phi^2\right)\simeq 0
\label{eq_sca_app}
\end{equation}
This expansion is equivalent to the assumed condition $\mu^2\gg \lambda \phi^2$. 
Substituting \eqref{fs} and \eqref{hs} and retaining terms to order ${\cal{O}}\left({\cal{C}} \frac{r^2}{r_s^2}\right)$, \eqref{eq_sca_app} becomes:
\begin{equation}
\phi''+\frac{2}{r}\left(1+\frac{3{\cal{C}}}{4}\frac{r2}{r_s^2}\right)\phi'+\left(1+2{\cal{C}}\frac{r^2}{r_s^2}\right)\mu^2\phi\simeq0.
\end{equation}
Using the boundary conditions $\phi(0)=\phi_c$ and $\phi'(0)=0$ we obtain the solution:
\begin{equation}
\phi(r)\simeq e^{-\alpha  \rho ^2} {}_1 F_1\left(\frac{1-\lambda }{2},\frac{3}{2},\beta ^2 \rho ^2\right)\phi_{c},\label{gr_sol}
\end{equation}
where
\begin{eqnarray}
\lambda&=&\frac{-\sqrt{{\cal{C}}} \sqrt{9 {\cal{C}}-32 \mu ^2 r_s^2}+9 {\cal{C}}+4 \mu ^2 r_s^2}{2 \sqrt{{\cal{C}}} \sqrt{9 {\cal{C}}-32 \mu ^2 r_s^2}},\label{lambda}\\
\alpha&=&\frac{1}{8} \left(\sqrt{{\cal{C}}} \sqrt{9 {\cal{C}}-32 \mu ^2 r_s^2}-3 {\cal{C}}\right),\label{alpha}\\
\beta&=&\frac{1}{2}{{\cal{C}}}^{1/4} {\left(9 {\cal{C}}-32 \mu ^2 r_s^2\right)^{1/4}},\label{beta}
\end{eqnarray}
$\rho=r/r_s$, and $_1 F_1$ is the Kummer confluent hypergeometric function. This solution is consistent with the behavior of the scalar field for $r\lesssim r_s$,  shown in Fig. \ref{fig_metrics_scalar}. Indeed, expanding \eqref{gr_sol} around $r=0$, we obtain $\phi(r) \simeq \phi_c \left(1-\frac{\mu^2 r^2}{6}\right)$, which corresponds to \eqref{eq_initial_symmetron3} under the assumption that $\mu^2 \gg \lambda \phi_c^{2}$. On the other hand, from \eqref{lambda}-\eqref{beta}, we derive that $\mu \lesssim \sqrt{\frac{9\,{\cal{C}}}{32 r_s^2}}$. For NS with ${\cal{C}}=0.25$, this condition translates to $\mu \lesssim 2.12\, r_0^{-1}$.

Close to the star's surface, the correction to the solution \eqref{gr_sol} scales as ${\cal{O}}({\cal{C}}^2)$. Hence, the solution \eqref{gr_sol} offers a reliable approximation to the precise profile of the scalar field within the star, with corrections typically around $4\%$ for NSs with ${\cal{C}}=0.25$.

We now proceed to quantify the back-reaction of the scalar field on the metric functions $\tilde{f}$ and $h$. As long as the conditions $\rho_c \gg \mu^2 \phi_c^{2}$ and $\kappa \gg \left |\xi\phi_c^2\right |$ hold true, $\tilde{f}$ and $h$ exhibit behavior analogous to GR. The former condition is consistently met, while the latter depends on the central density. In practice, neglecting this term results in an error below ${\cal{O}}(0.1)$ for $\xi \lesssim 0.1$, across all central density values. However, increasing $\xi$ beyond $0.1$ leads to significant error amplification, exceeding $20\%$. In particular, $\xi\phi_c^2$ becomes important for $\xi \gtrsim 0.2$ in the low-density regime. Consequently, for coupling $\xi\gtrsim 0.1$, the Schwarzschild interior solutions are no longer a satisfactory approximation to \eqref{eq_initial_symmetron1}-\eqref{eq_initial_symmetron3}.

\section{Dimensionless form of equations}\label{adim_eqs}
In this appendix, we present Einstein's field equations, the scalar field equation, and the equation for angular velocity in dimensionless forms.

The dimensionless scalar of torsion and the boundary term are expressed in terms of the variable $s$ as follows:
\begin{eqnarray}
	T(s)&=&-\frac{\left(h-1\right)\left(-\tilde{f}+\tilde{f} h -2\tilde{f}_{,s}\right)}{4 \pi \tilde{f} h^{2}}e^{-2s} \, , \\
	B(s)&=&\frac{2\tilde{f}_{,s}\tilde{f}\left(h_{,s}+2h^{2}-4h\right)+h\left(-\tilde{f}_{,ss}+2\tilde{f}_{,s}^{2}+\tilde{f}_{,s}\right)+4\tilde{f}^{2}\left(h_{,s}+h^{2}-h\right)}{8\pi \tilde{f}^{2}h^{3}}e^{-2s} \, .
\end{eqnarray}
The modified Tolman-Oppenheimer-Volkoff (TOV) equations (\ref{eq_tov1}), (\ref{eq_tov2}), (\ref{eq_mass})  and (\ref{eq_tov4}), and the scalar field equation \eqref{eq_tov3} become:
\begin{eqnarray}
	\frac{\tilde{f}_{,s}}{\tilde{f}}&=&\frac{h^{2}-1}{2}+\frac{\pi h^{2}\left(4P+2\tilde{\mu}^{2}\tilde{\phi}^{2}-\tilde{\lambda} \tilde{\phi}^{4}\right)}{1+\xi \tilde{\phi}^{2}}e^{2s}+\frac{\tilde{\phi}_{,s}^{2}}{4\left(1+\xi \tilde{\phi}^{2}\right)}, \\
	\frac{h_{,s}}{h}&=& \frac{1-h^{2}}{2}+\frac{\pi h^{2}\left(4\rho-2\tilde{\mu}^{2}\tilde{\phi}^{2}+\tilde{\lambda} \tilde{\phi}^{4}\right)}{1+\xi \tilde{\phi}^{2}}e^{2s}-\frac{2\xi \left(h-1\right)\tilde{\phi}_{,s}\tilde{\phi}}{1+\xi \tilde{\phi}^{2}}+\frac{\tilde{\phi}_{,s}^{2}}{4\left(1+\xi \tilde{\phi}^{2}\right)}, \\
	m_{,s}&=&\frac{ \left(4 \pi   h ^2 \left(4 \rho -2 \mu^2 \tilde{\phi} ^2+\lambda \tilde{\phi} ^4\right)e^{2 s}+\tilde{\phi}_{,s} (8 \xi \tilde{\phi} +\tilde{\phi}_{,s})-8 \xi h  \tilde{\phi}  \tilde{\phi}_{,s}\right)}{16 h ^2 \pi \left(1 + \xi \tilde{\phi} ^2 \right)}3 e^s,\\
\rho_{,s}&=&\frac{\left(\rho+P\right)}{4P'(\rho)}\left[2\left(1-h^{2}\right)-\frac{4\pi h^{2}\left(4P+2\tilde{\mu}^{2}\tilde{\phi}^{2}-\tilde{\lambda} \tilde{\phi}^{4}\right)e^{2s}+\tilde{\phi}_{,s}^{2}}{1+\xi \tilde{\phi}^{2}}\right],
\end{eqnarray}
\begin{eqnarray}
\tilde{\phi}_{,ss}&=&\frac{1}{1+\xi \tilde{\phi}^2}\Bigg\{2\xi h \tilde{\phi}((h-1)^{2}(1+\xi \tilde{\phi}^2)+2\pi e^{2s}h(h-1)(4P+2\tilde{\mu}^{2}\tilde{\phi}^{2}-\tilde{\lambda} \tilde{\phi}^{4})\nonumber \\
	&-& 4\pi e^{2s}h\tilde{\phi}^{2}(\tilde{\mu}^{2}-\tilde{\lambda} \tilde{\phi}^{2})) -h^{2}\tilde{\phi}_{,s}(1+\xi \tilde{\phi}^{2}-2 \pi e^{2s}(2P-2\rho+ 2\tilde{\mu}^{2}\tilde{\phi}^{2}-\tilde{\lambda} \tilde{\phi}^{4}))\nonumber \\
	&-&8\pi e^{2s} h^{2} \tilde{\phi}(\tilde{\mu}^{2}-\tilde{\lambda} \tilde{\phi}^{2})-(h-1)\xi \tilde{\phi} \tilde{\phi}_{,s}\Bigg\}.
	\end{eqnarray}
The equation for the angular velocity, Eq.(\ref{eq_tov5}), is:
\begin{eqnarray}
	\bar{\omega}_{ss}&-&\frac{1}{2\left(1+\xi \tilde{\phi}^{2}\right)}\Bigg\{\left[8\pi e^{2s}h^{2}\left(\rho+P\right)-4\xi h \tilde{\phi}_{,s}\tilde{\phi}-6\left(2\xi \tilde{\phi}_{,s}\tilde{\phi}+\xi \tilde{\phi}^{2}+1\right)+\tilde{\phi}_{,s}^{2}\right]\bar{\omega}_{s} \nonumber \\ &-& 32h\left[\pi e^{2s}h\left(\rho+P\right)-\xi \tilde{\phi}_{,s}\tilde{\phi}\right]\bar{\omega}+32\xi h \tilde{\phi}_{,s}\tilde{\phi}\Bigg\}=0 \, .
\end{eqnarray}
The initial conditions (\ref{eq_initial_symmetron1})-(\ref{eq_initial_symmetron5}) and  \eqref{eq_initial_omega} are given by:

\begin{eqnarray}
	\tilde{f}(s)&\simeq&1+\frac{2 \pi \left(6P_{c}+2\rho_{c}+2\tilde{\mu}^{2}\tilde{\phi}_{c}^{2}-\tilde{\lambda} \tilde{\phi}_{c}^{4}\right)}{3\left(1 +\xi \tilde{\phi}_{c}^{2}\right)} e^{2s_{0}},\\
	h(s) &\simeq&h_{c}+\frac{\pi \left(4\rho_{c}-2\tilde{\mu}^{2}\tilde{\phi}_{c}^{2}+\tilde{\lambda} \tilde{\phi}_{c}^{4}\right)}{3\left(1 +\xi \tilde{\phi}_{c}^{2}\right)} e^{2s_{0}}, \\
	\tilde{\phi}(s)&=&\tilde{\phi}_{c}\left[1-\frac{4}{3}\pi \left(\tilde{\mu}^{2}-\tilde{\lambda} \tilde{\phi}_{c}^{2}\right)\right]e^{2s_{0}}, \\
	P(s)&\simeq & P_{c}+\frac{\pi \left(\rho_{c}+P_{c}\right)\left(6P_{c}+2\rho_{c}+2\tilde{\mu}^{2}\tilde{\phi}_{c}^{2}-\tilde{\lambda} \tilde{\phi}_{c}^{4}\right)}{3\left(1+\xi \tilde{\phi}_{c}^{2}\right)}e^{2s_{0}},\\
m(s)&\simeq&\frac{3 e^{3 s_0} \left(4 \rho_c-2\mu^2 \tilde{\phi}_c^2 +\lambda\tilde{\phi}_c^4\right)}{8(1+ \xi \tilde{\phi}_c^2)},\\
\bar{\omega}(s)&\simeq&\bar{\omega}_{c}+\frac{16\pi e^{2s_0}}{15\left(1+\xi \tilde{\phi}_{c}^{2}\right)}\left[\xi \tilde{\phi}_{c}^{2}\left(\tilde{\mu}^{2}-\tilde{\lambda} \tilde{\phi}_{c}^{2}\right)\left(\bar{\omega}_{c}-\Omega\right)+3\bar{\omega}_{c}\left(\rho_{c}+P_{c}\right)\right] \, .
\end{eqnarray}

 \acknowledgments
 We thank the referee for fruitful suggestions regarding the manuscript. KN and HB thank the Algerian Ministry of Higher Education and Scientific Research (MESRS) for financial support.

\bibliographystyle{utphys}
\bibliography{references}
	
\end{document}